\documentclass[twocolumn]{aastex63}

\usepackage{times}

\usepackage{graphicx}

\usepackage{tikz}

\usepackage{listings}
\usepackage[T1]{fontenc}
\usepackage{aecompl}
\usepackage{calligra}
\DeclareMathAlphabet{\mathcalligra}{T1}{calligra}{m}{n}
\DeclareFontShape{T1}{calligra}{m}{n}{<->s*[2.2]callig15}{}

\usepackage{hyperref}
\hypersetup{
    pdfnewwindow=true,      % links in new window
    colorlinks=true,       % false: boxed links; true: colored links
    linkcolor=orange,      % color of internal links
    citecolor=cyan,        % color of links to bibliography
    filecolor=orange,      % color of file links
    urlcolor=orange           % color of external links
}

\def\apjl{ApJL}               % Astrophysical Journal, Letters
\def\apjs{ApJS}               % Astrophysical Journal, Supplement
\def\aap{A\&A}                % Astronomy and Astrophysics
\defcitealias{DOrazioDuffell:2021}{DD21}
\defcitealias{TiedeDOrazio:2023}{TD23}

\usepackage{color}

\def\Fit{\rm{Rec}}

\def\min{\rm{min}}

\def\Msun{{M_{\odot}}}
\def\Mdot{\dot{M}}
\def\Mag{\mathcal{M}}

\def\eg{{\em e.g.}}
\def\ie{{\em i.e.}}

\def\codename{\texttt{binlite}}

\begin{document}

\shorttitle{Eccentric Binary Accretion Rates}
\shortauthors{D'Orazio et al.}

\title{Fast Methods for Computing Photometric Variability of Eccentric Binaries:\\
Boosting, Lensing, and Variable Accretion}

\author[0000-0002-1271-6247]{Daniel J. D'Orazio}
\affiliation{Niels Bohr International Academy, Niels Bohr Institute, Blegdamsvej 17, 2100 Copenhagen, Denmark}
\email{daniel.dorazio@nbi.ku.dk}

\author[0000-0001-7626-9629]{Paul C. Duffell}
\affiliation{Department of Physics and Astronomy, Purdue University, 525 Northwestern Avenue, West Lafayette, IN 47907-2036, USA}
\email{pduffell@purdue.edu}

\author[0000-0002-3820-2404]{Christopher Tiede}
\affiliation{Niels Bohr International Academy, Niels Bohr Institute, Blegdamsvej 17, 2100 Copenhagen, Denmark}
\email{christopher.tiede@nbi.ku.dk}
%%%%%%%%%%%%%%%%%%%%%%%%%%%%%%%%%%%%%%%%%%%%%%%%%%%%%%%%%%%%%%%%
%%%%%%%%%%%%%%%%%%%%%%%%%%%%%%%%%%%%%%%%%%%%%%%%%%%%%%%%%%%%%%%%

\begin{abstract} 
We analyze accretion-rate time series for equal-mass binaries in co-planar gaseous disks spanning a continuous range of orbital eccentricities up to $0.8$, for both prograde and retrograde systems. The dominant variability timescales match that of previous investigations; the binary orbital period is dominant for prograde binaries with $e \gtrsim 0.1$, with a $5\times$ longer ``lump'' period taking over for $e\lesssim0.1$. This lump period fades and drops from  $5\times$ to $4.5\times$ the binary period as $e$ approaches $0.1$, where it vanishes. For retrograde orbits, the binary orbital period dominates at $e\lesssim0.55$ and is accompanied by a $2\times$ longer-timescale periodicity at higher eccentricities. The shape of the accretion-rate time series varies with binary eccentricity. For prograde systems, the orientation of an eccentric disk causes periodic trading of accretion between the binary components in a ratio that we report as a function of binary eccentricity. We present a publicly available tool, \codename, that can rapidly ($\lesssim0.01$~sec) generate templates for the accretion-rate time series, onto either binary component, for choice of binary eccentricity below $0.8$. As an example use-case, we build lightcurve models where the accretion rate through the circumbinary disk and onto each binary component sets contributions to the emitted specific flux. We combine these rest-frame, accretion-variability lightcurves with observer-dependent Doppler boosting and binary self-lensing. This allows a flexible approach to generating lightcurves over a wide range of binary and observer parameter space. We envision \codename~as the access point to a living database that will be updated with state-of-the-art hydrodynamical calculations as they advance. 
\end{abstract}


\section{Introduction}
The binary-gas-disk interaction arises in a multitude of astrophysical environments. It is important for binary orbital evolution \citep[see, \eg,][]{LaiMunoz:Review:2022}: from sculpting planetary system architectures \citep[\eg,][]{Ward:1997, NelsonBook:2018}, to impacting stellar binary demographics \citep[\eg,][]{Valli+2024}, to facilitating mergers of stellar-mass \citep[\eg,][]{Stone_AGNchan_2017} and supermassive \citep[\eg,][]{Begel:Blan:Rees:1980} black hole binaries (SMBHBs).  

Gas disks also offer a way to observe such systems in the electromagnetic (EM) spectrum through accretion. In the realm of stellar binary+disk systems, radio and millimeter-wavelength observations have revealed a variety of disks feeding and forming young stellar binaries as well as planetary systems \citep[\eg,][]{Tobin+2016, Alves+2019, Czekala_CBD_ALMA+2021, Zurlo_imgCBDrev_2023}. In the time domain, photometric variability associated with periodic accretion onto stellar binaries has been observed in multiple systems \citep[\eg,][]{Tofflemire_DQTau+2017, Tofflemire_TWA3A+2017}, providing further data on the disk+binary interaction.

Resolving disks around SMBHBs is more difficult than in the stellar case due to a lack of known sub-parsec-separation SMBHBs, uncertainty in the emission structure surrounding the accreting SMBHB at the relevant wavelengths, and the high spatial resolution needed for imaging of these distant sources \citep[see][]{DOrazioLoebVLBI:2018, DOrazioLoeb_Gaia:2019}.
Similar to the stellar analogue, however, accretion onto SMBHBs could be observable via bright, periodically modulated EM emission. 
While there is no definitive evidence for the sub-parsec-separation SMBHBs that will merge within the age of the Universe, accretion rates onto these systems can be as high as for single SMBHs \citep[\eg,][]{FarrisGold:2012, DHM:2013:MNRAS}, suggesting that they could be a sub-population of the quasars. In addition to being bright, such a binary-quasar population could be identified by its periodic imprint on quasar lightcurves, on $\sim$year or shorter timescales \citep{HKM09, KelleyLens+2021, XinHaiman_LSSTshort:2021, RomanPLCs+2023}. Such periodicity can arise from the binary's modulation of the accretion rate \citep[\eg,][]{ Hayasaki:2008, MacFadyen:2008, DHM:2013:MNRAS, Farris:2014}, or observer-dependent relativistic effects due to the binary orbit \citep{PG1302Nature:2015b, DoDi:2018, Spikey:2020}. Both offer a way to identify such systems in photometric time-domain data, with multiple searches having identified $\sim250$ candidates to date \citep[see][and references therein]{DOrazioCharisi:2023}.

Time-domain searches require predictions for periodic signatures of binary accretion and also characterization of the intrinsic variability noise \citep[\eg,][for the SMBHB case]{Vaughan2016, ZhuThrane:2020}.  Here we make a step towards the former, by characterising the variable accretion rates of eccentric binaries embedded in circumbinary disks (CBDs). We present an analysis of accretion variability measured from 2D isothermal numerical hydrodynamical calculations of gas disks accreting onto equal mass binaries, for a continuous range of binary eccentricities $e\leq0.8$ and for both prograde \citet[][hereafter \citetalias{DOrazioDuffell:2021}]{DOrazioDuffell:2021} and retrograde configurations of the binary and disk angular momentum \citet[][hereafter \citetalias{TiedeDOrazio:2023}]{TiedeDOrazio:2023}. We use this data to build a publicly available tool named \codename~that can rapidly generate accretion-rate time series data, for any binary eccentricity in the simulated range via Fourier decompositions of the simulation data. 

Section \ref{S:Method} describes our methods while Sections \ref{S:Results_2DPow}, \ref{Ss:Results:Mdotcurves}, and \ref{S:Qrat} present the results of our periodicity analysis, accretion-rate time series reconstruction, and calculation of preferential accretion rates. As an example use-case and to demonstrate the wide range of periodic lightcurves that can arise from accreting eccentric binaries, Section \ref{S:Application} presents a method for generating light-curves of accreting black hole binaries, in a chosen observing band, while including the observer-dependent relativistic effects of Doppler boosting and gravitational self-lensing for multiple observer viewing angles.

We envision this tool and the data it is built from as a starting point from which further sophistication in numerical models and post-processing 
can be added with the goal of generating a publicly available, living-lightcurve database for modelling, interpreting, and searching for emission from accreting binary systems. We discuss these future prospects and current limitations in Section \ref{S:Conclusion}.

\section{Methods}
\label{S:Method}
Throughout we consider a binary of total mass $M$, with equal mass components (described by mass ratio $q\equiv M_2/M_1=1$), orbital eccentricity $e$, semi-major axis $a$, and orbital angular frequency $\Omega_b$. A locally isothermal, circumbinary disk accretes onto the binary and is modelled with viscous hydrodynamics in the two dimensions in the plane of the binary orbit. In this case the disk is characterized by the disk aspect ratio in vertical hydrostatic equilibrium, $h$, which describes the relative importance of pressure forces, and the kinematic coefficient of viscosity $\nu$. Because of the simplified physics, one can scale results to any value of $M$ or $a$, which amounts to choosing an orbital timescale via $\Omega_b$. Throughout we scale the accretion rate by the equivalent steady-state value for a single mass, $\dot{M}_0 = 3 \pi \Sigma_0 \nu$, for arbitrary surface-density scale $\Sigma_0$. We consider both prograde and retrograde configurations of the binary orbit with respect to the CBD. In both cases the equations of hydrodynamics are solved using the moving-mesh code \texttt{DISCO} \citep{DuffellMHDDISCO:2016}.

\paragraph{Accretion From a Prograde Circumbinary Disk}
The accretion-rate time-series data for prograde disks around eccentric binaries is taken directly from the output of 2D numerical viscous hydrodynamical calculations described in \citetalias{DOrazioDuffell:2021}. These calculations assume a locally isothermal equation of state, which keeps the aspect-ratio a constant value of $h=0.1$ in the circumbinary disk, and a spatially constant coefficient of kinematic viscosity $\nu=10^{-3}a^2 \Omega_b$. Specifically, we utilize the main calculations in \citetalias{DOrazioDuffell:2021}, which evolve the binary and disk for 25,000 binary orbits, with the first 500 orbits relaxing the disk around a binary on a circular orbit, and the following 20,000 binary orbits sweeping the binary eccentricity linearly from $e=0$ to $e=0.9$. Here we utilize the accretion rates measured via a sink prescription (Eq. 3 of \citetalias{DOrazioDuffell:2021}) onto each component of the binary, as a function of time.

\paragraph{Accretion From a Retrograde Circumbinary Disk}
The accretion-rate time-series data for retrograde disks around eccentric binaries is taken directly from the output of 2D numerical viscous hydrodynamical calculations described in \citetalias{TiedeDOrazio:2023}. These calculations assume the same disk and binary parameters as the prograde case except that the binary eccentricity is swept linearly from $e=0.0-0.8$ over a timescale of 10,500 orbits, with 500 orbits to relax the disk around a binary with a circular orbit. We note that the accretion sinks in \citetalias{TiedeDOrazio:2023} have half the characteristic sink size and are implemented to be ``torque free'', compared to the standard sink implementation in \citetalias{DOrazioDuffell:2021}  \citep[see][for further clarificaiton on these sink types]{Dempsey-TFsinks:2020, DittmannRyan:2021}.

\subsection{Accretion-Rate Variability Timescales}
We first compute the dominant variability timescales as a function of binary orbital eccentricity. We follow \citet{Duffell:2020}; \citetalias{TiedeDOrazio:2023} and compute a 2D periodogram of the accretion-rate time series by taking the norm of the quantity, 
\begin{equation}
    \mathcal{P}(e, \omega) =  \frac{1}{\sqrt{2 \pi \sigma^2_\mathcal{P}}} \int^{t(e_f)}_{t(e_0)}{  \mathrm{e}^{-\frac{1}{2}\frac{\left(t(e)-\tau\right)^2}{\sigma^2_\mathcal{P}}}  \Mdot(\tau) \mathrm{e}^{-i \omega \tau} d\tau},
    \label{eq:2Dpdgm}
\end{equation}
which picks out the power in Fourier components with frequency $\omega$ in a window of the accretion-rate time series centered on time $t(e)$ and of characteristic width $2\sqrt{2 \log2} \sigma_\mathcal{P}$. For our choice of $\sigma_\mathcal{P} = 30 (2 \pi \Omega^{-1}_b)$, this corresponds to a small window of $\sim 70$ orbits in eccentricity centered around any $e$ in the time series.

%%%%%%%%%%%%%%%%%%%%%%%%%%%%%%%%%%%%%%%%%%%%%%%%
%%% 2D FT %%%
%%%%%%%%%%%%%%%%%%%%%%%%%%%%%%%%%%%%%%%%%%%%%%%%
\begin{figure*}
\begin{center}$
\begin{array}{cc}
\hspace{-15pt}
\includegraphics[scale=0.6]{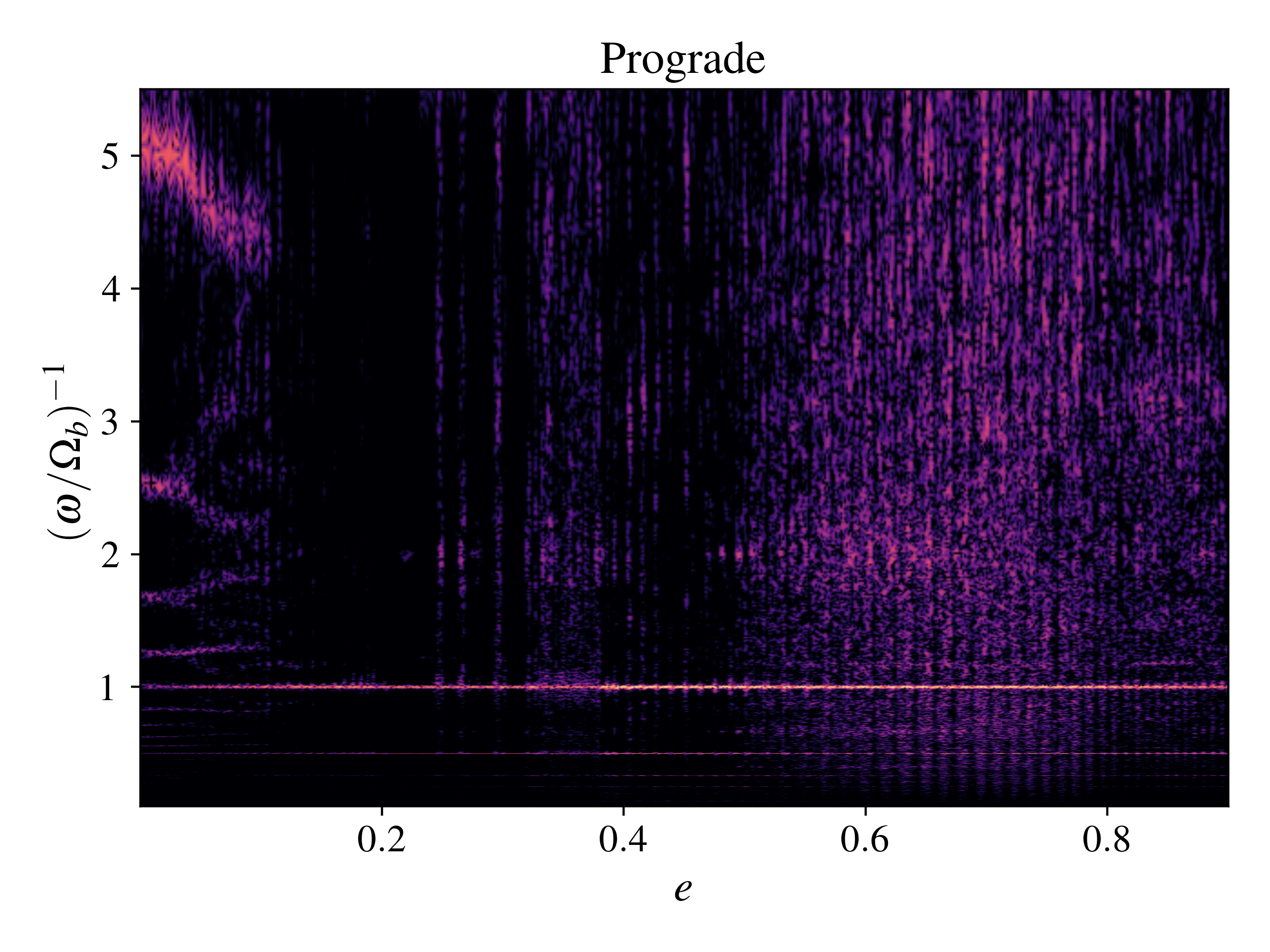}  & \hspace{-30pt}
\includegraphics[scale=0.6]{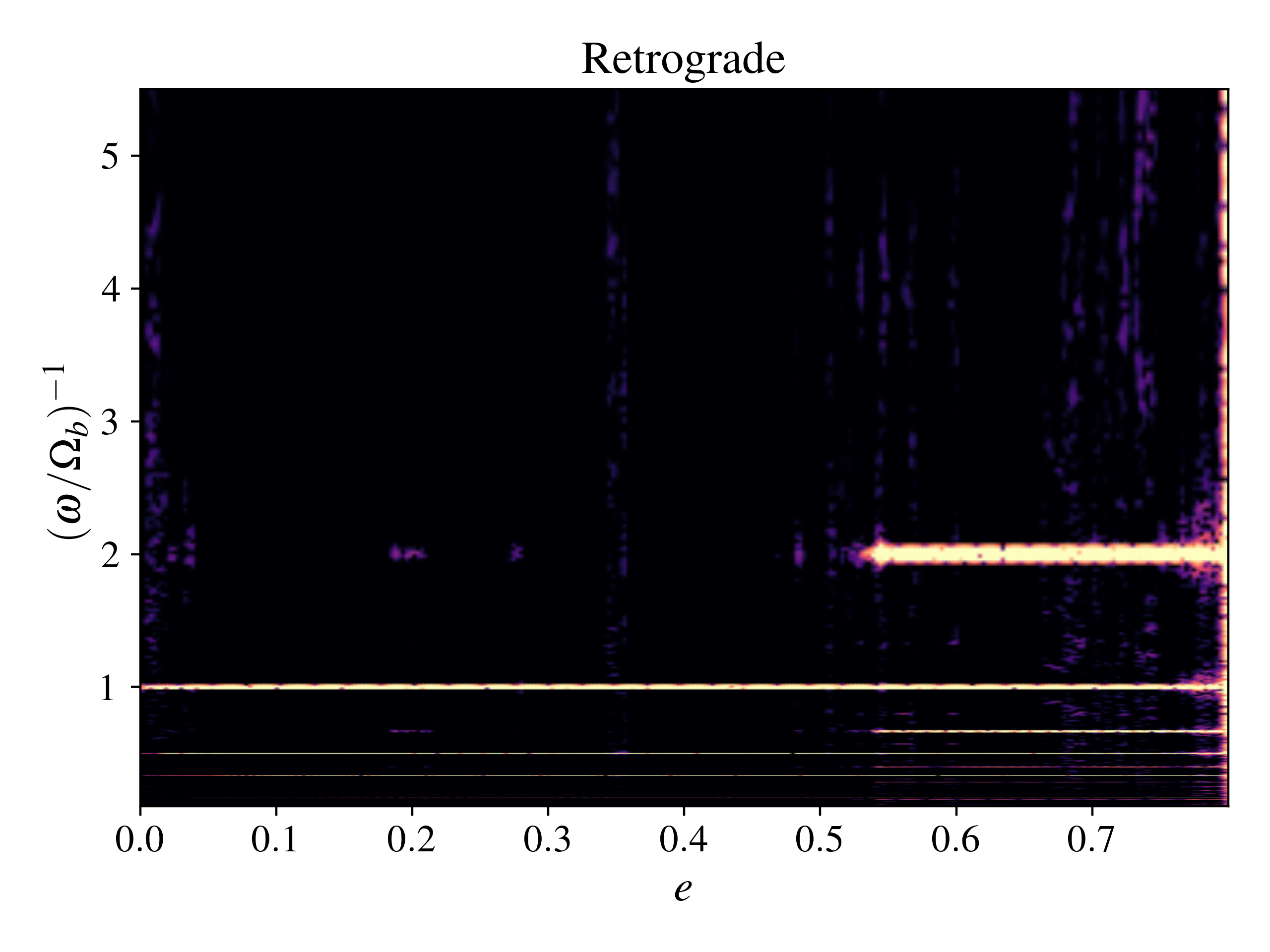} 
\end{array}$
\end{center}
\vspace{-20pt}
\caption{
 2D periodograms of the total accretion rate from prograde (left) and retrograde (right) circumbinary disks onto binaries with orbital eccentricity $e$ ranging from 0.0 to 0.8. The y-axis indicates the timescale in units of orbital periods. Black indicates regions where there is no power, while purple and yellow regions have increasingly more power. Both panels are normalized to the same color scale.
}
\label{Fig:2DPow}
\end{figure*}
%%%%%%%%%%%%%%%%%%%%%%%%%%%%%%%%%%%%%%%%%%%%%%%%

\subsection{Fourier Reconstruction}
\label{Ss:Fourier_Reconstruction}
Our primary goal is to generate accretion-rate time series, which are periodic over the binary orbital period, for any chosen value of the orbital eccentricity. This is possible given the continuous sweep of our solutions through binary parameter space.
We start with the total accretion rate onto the binary $\Mdot(t)$ and its two components $\Mdot_1(t)$ and $\Mdot_2(t)$ computed over the entire 25,000 (10,500) orbit sweep from the numerical calculation of \citetalias{DOrazioDuffell:2021} (\citetalias{TiedeDOrazio:2023}). Because these calculations carried out a linear sweep in eccentricity with time,  we also have a linear relation between the time $t$ and the orbital eccentricity, $t(e)$.

The accretion-rate time series at a given orbital eccentricity, onto either component of the binary, is reconstructed with a Fourier series,
\begin{eqnarray}
\Mdot_{\Fit}(t,e) = \alpha_{0} &+& \sum_n \alpha_{n}(e) \cos{\left(n \Omega t\right)} \nonumber \\
&+& \sum_n \beta_{n}(e) \sin{\left(n \Omega t\right)},
\label{Eq:F_rec}
\end{eqnarray}
for chosen fundamental frequency $\Omega$ and its integer multiples.
That is, a simple, yet accurate reconstruction is possible when the primary power is concentrated at integer multiples of one frequency. A natural choice is the binary frequency $\Omega_b$, and we show in Section~\ref{S:Results_2DPow} that this is indeed the best choice except for a few regions of parameter space where a lower frequency dominates, but still at approximate integer multiples of $\Omega_b$. 

The Fourier amplitudes are computed for the nearly continuous range of orbital eccentricities by convolving the accretion-rate time series with a Gaussian centered around a chosen eccentricity $e$ that arises at time $t(e)$ in the eccentricity sweep,
\begin{eqnarray}
\label{Eq:F_amps}
\alpha_0(e) &=& \frac{1}{\sqrt{2 \pi \sigma^2}} \int^{t(e_f)}_{t(e_0)} 
            \Mdot(t') 
            e^{-\frac{1}{2}\left[ \frac{ \left(t' - t(e)\right)^2 }{\sigma^2}\right]}  \ dt'  \\
\alpha_n(e) &=& \frac{2}{\sqrt{2 \pi \sigma^2}} \int^{t(e_f)}_{t(e_0)} 
            \Mdot(t') 
            e^{-\frac{1}{2}\left[ \frac{ \left(t' - t(e)\right)^2 }{\sigma^2}\right]} \cos\left({n \Omega t'}\right) \ dt' \nonumber  \\
\beta_n(e) &=& \frac{2}{\sqrt{2 \pi \sigma^2}} \int^{t(e_f)}_{t(e_0)}
            \Mdot(t') 
            e^{-\frac{1}{2}\left[ \frac{ \left(t' - t(e)\right)^2 }{\sigma^2}\right]} \sin\left({n \Omega t'}\right) \ dt', \nonumber
\end{eqnarray}
where $\sigma$ chooses the window width of $\Mdot$ times-series data over which to construct the Fourier series. In practice, we calculate reconstructions up to $e=0.81$ for prograde systems and $e=0.791$ for retrograde systems. Throughout we choose $\sigma = 10 (2 \pi \Omega^{-1}_b)$, which corresponds to a Gaussian full width at half maximum of $\approx23.5$ orbits.

Tracking accretion rates onto both components of an equal-mass-ratio binary is necessary because, for prograde disks, some binary eccentricities excite disk eccentricities that allow the accretion-rate to periodically favor one binary component over the other \citep{Dunhill+2015, MunozLai_PulsedEccAcc:2016, Siwek+2023}.
The simulations of \citetalias{DOrazioDuffell:2021} find that for $0.0 \leq e\leq 0.18$ and $e\geq0.38$ the cavity is eccentric and precesses on super-orbital timescales ($\mathcal{O}(10^2)$ orbital periods). This causes the accretion to favor one binary component for approximately one half of the precession period of the eccentric cavity. When binary and cavity eccentricity vectors pass through a perpendicular configuration, the accretion-rate ratio quickly swaps to favor the other component for the other half of the cavity precession period. 
Hence, when the circumbinary cavity is eccentric and precessing, for prograde disks around binaries with eccentricities in the range $0.0 \leq e\leq 0.18$ and $e\geq0.38$, there are three possible accretion states: one where the primary dominates accretion, one where the secondary dominates accretion, and one shorter-lived stated where the two share the accretion rate as they swap between the first two states. In these cases the accretion-rate ratio averages to unity when taken over a disk precession period.

\citetalias{DOrazioDuffell:2021} finds that prograde disks around binaries with eccentricities in the range $0.18 \leq e \leq 0.38$ are much more symmetric around the origin and either do not precess (due to lack of disk eccentricity or to locking with the binary eccentricity vector) or have much longer precession periods than for higher or lower binary eccentricities. Similar observations were made from the numerical calculations of \citet{MirandaLai+2017} and \citet{Siwek+2023}. \citet{Siwek+2023} classify regions of parameter space where the disk eccentricity is ``locked'' to the binary eccentricity, finding such a state for $e=0.2$, $q=1$, and otherwise finding precessing states for $e=0.0,0.4,0.6,0.8$, in agreement with \citetalias{DOrazioDuffell:2021}. 
Here we find that even in the locked regime, the small asymmetry of the disk can still cause the accretion rate to be spilt unequally between the binary components, but with a different nature than for the precessing solutions. While the eccentric binary-disk dynamics are worth understanding further in this regime, for the purposes of this study, we note that even in this ``symmetric" non-precessing disk state, asymmetries arise that cause unequal accretion rates onto the binary components (See Section~\ref{S:Application}).

For retrograde systems, where persistent disk eccentricities are not excited, the accretion rates are always split evenly between the binary components.

%%%%%%%%%%%%%%%%%%%%%%%%%%%%%%%%%%%%%%%%%%%%%%%%
%%% Prog Mdot Sim and Rec %%%
%%%%%%%%%%%%%%%%%%%%%%%%%%%%%%%%%%%%%%%%%%%%%%%%
\begin{figure*}
\vspace{-10pt}
\begin{center}$
\begin{array}{cc}
\includegraphics[scale=0.35]{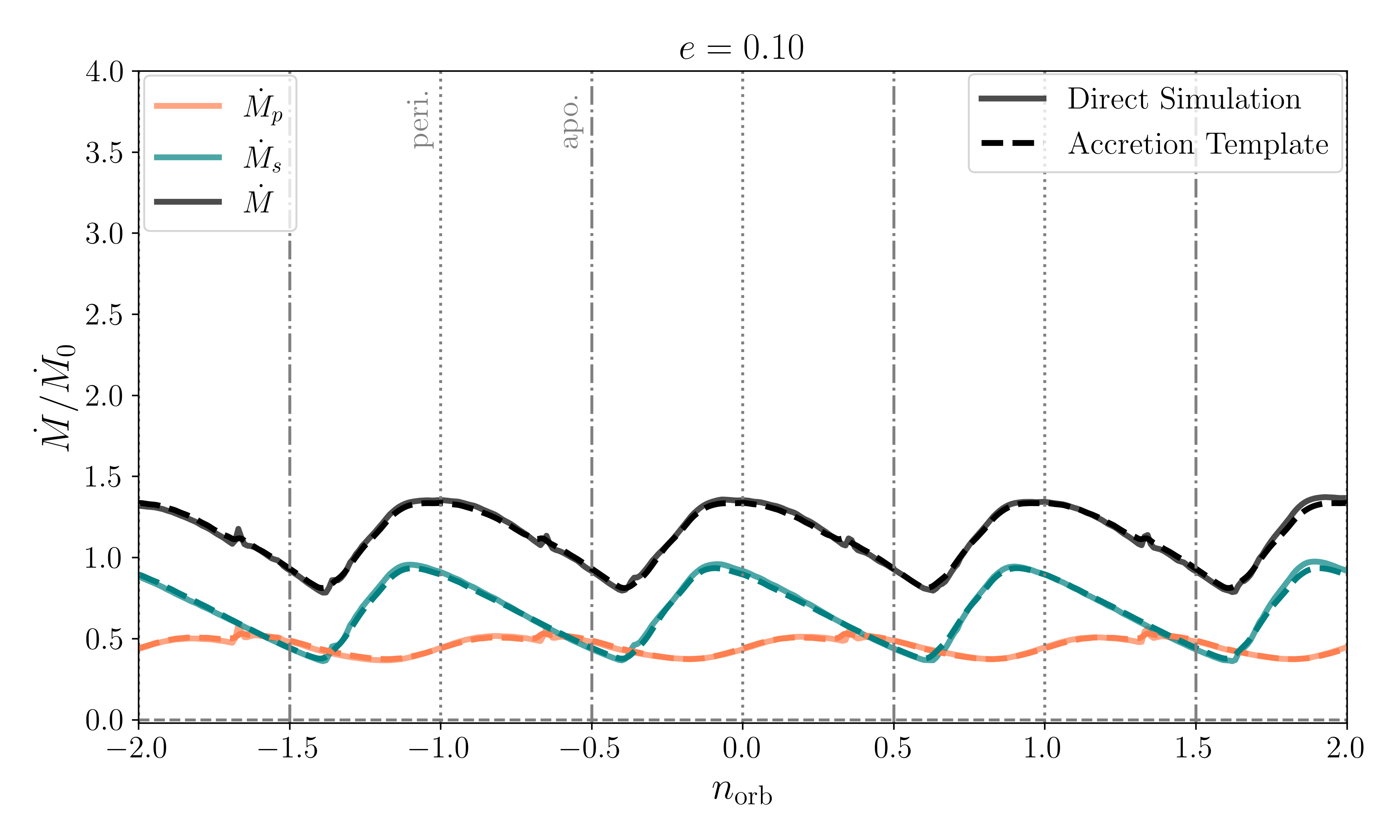} &
\includegraphics[scale=0.35]{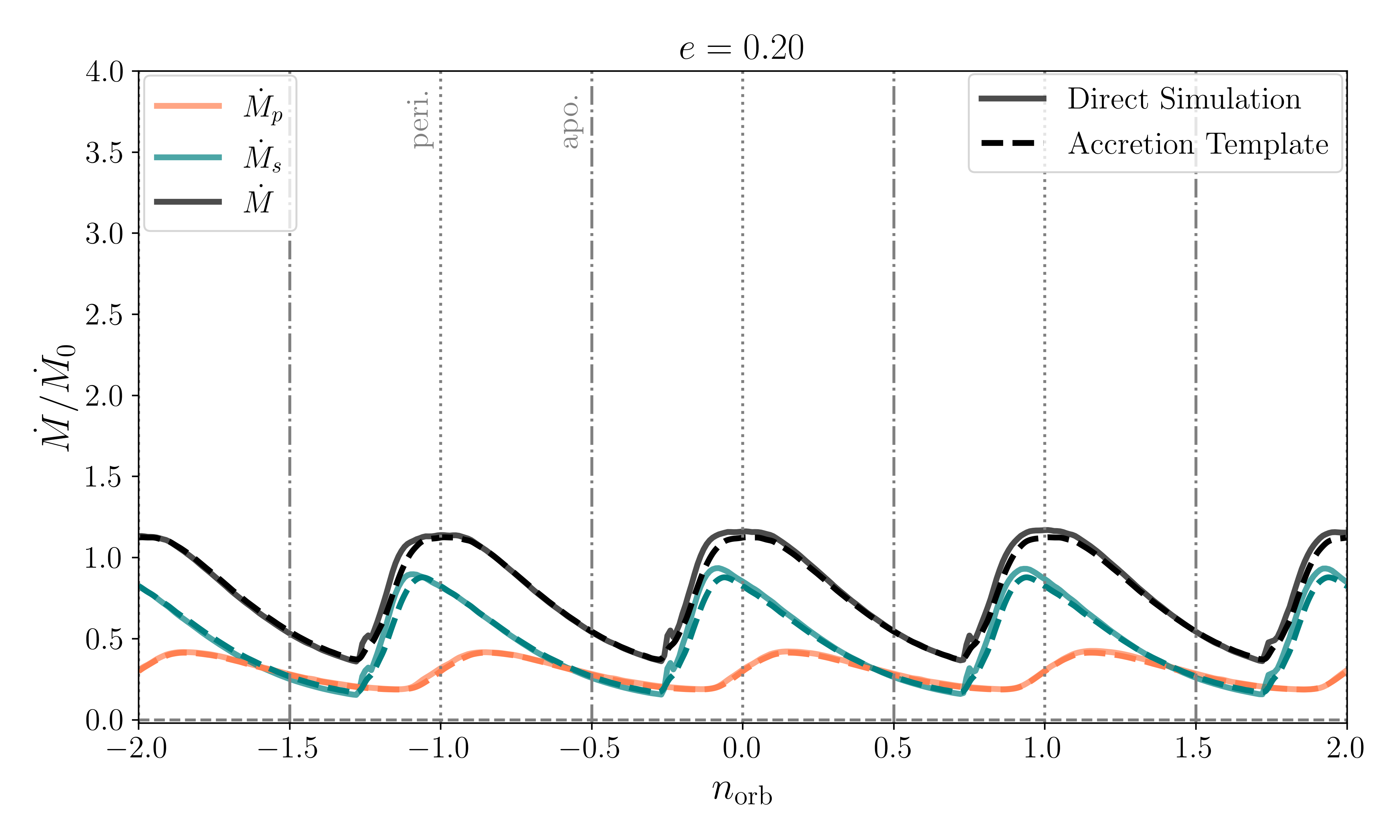} \vspace{-23pt} \\
\includegraphics[scale=0.35]{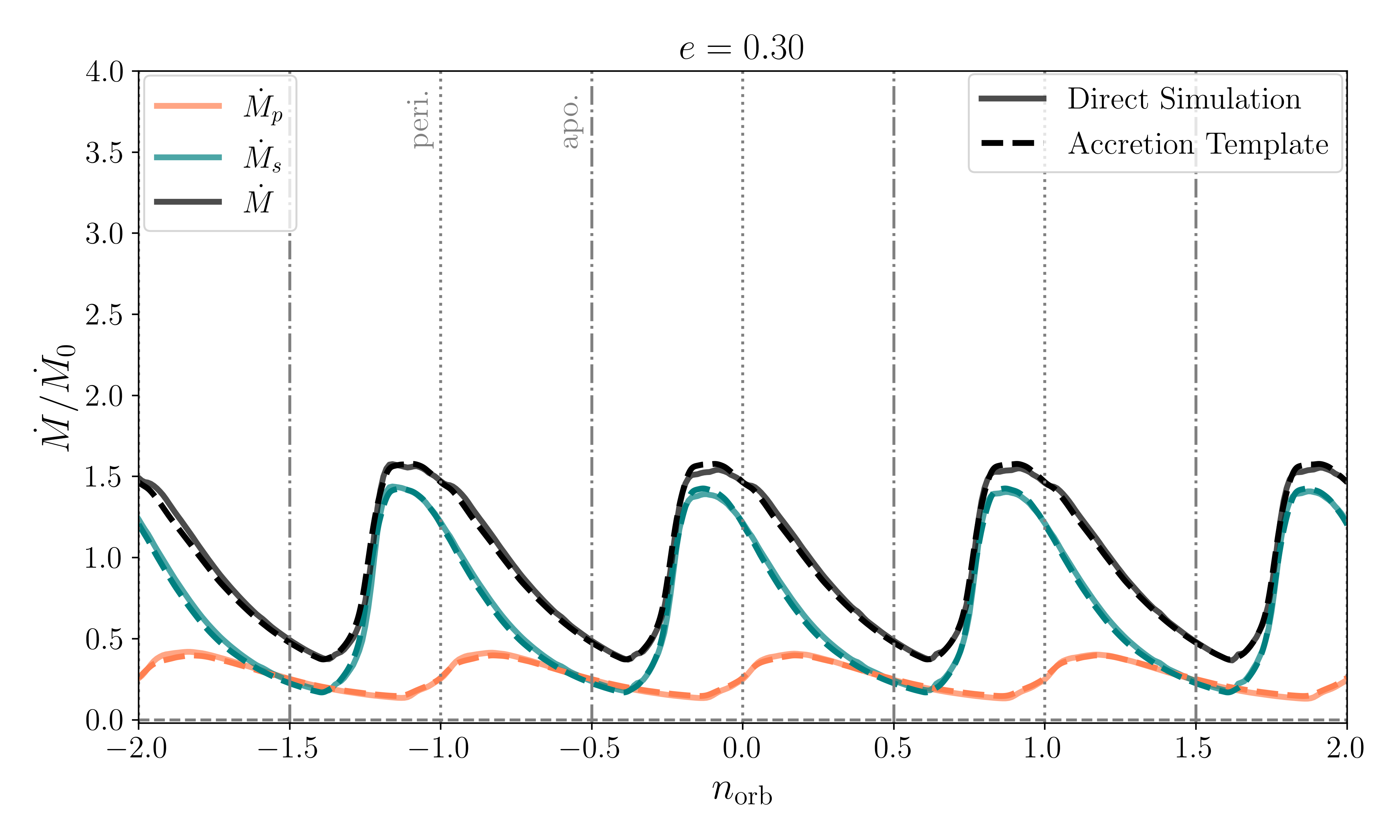} & 
\includegraphics[scale=0.35]{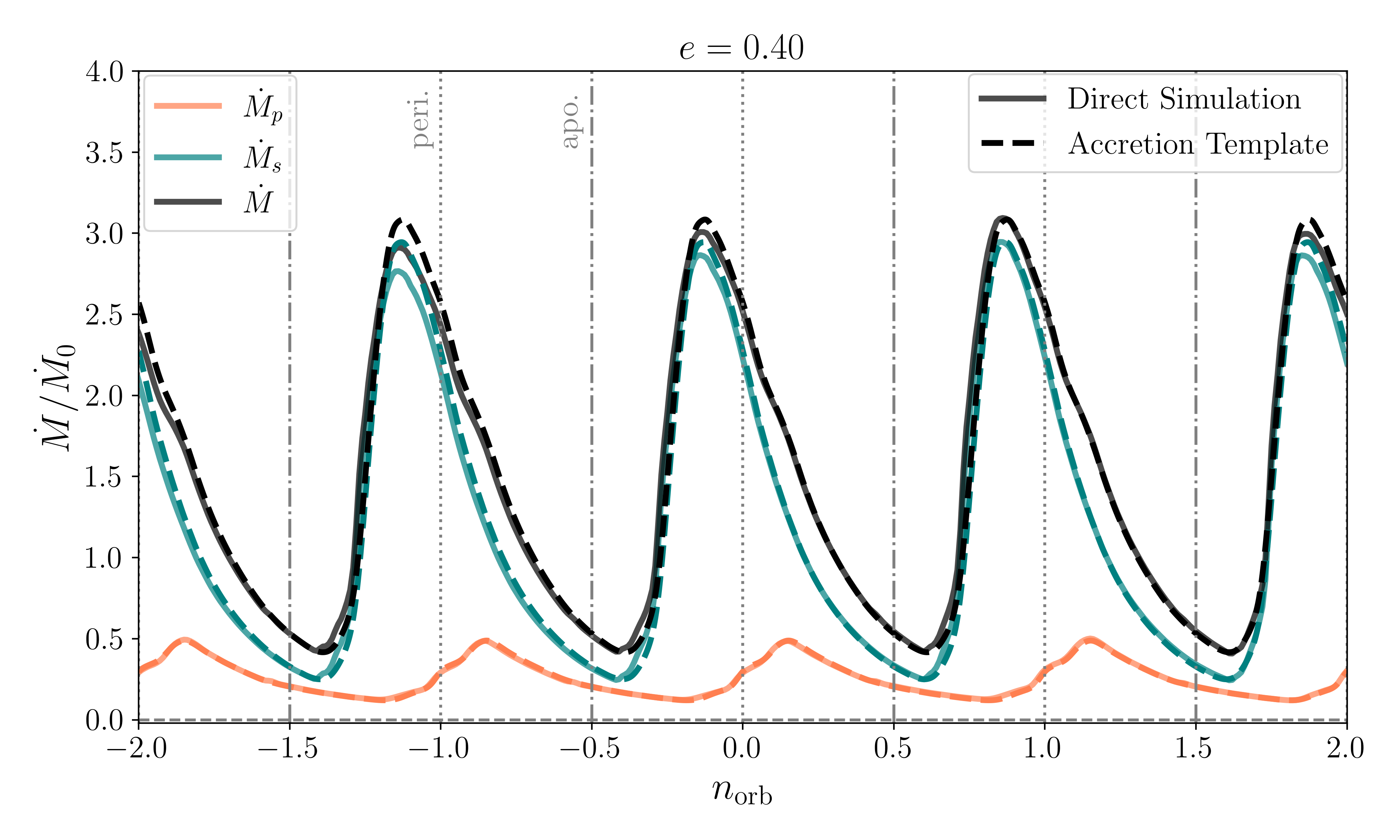} \vspace{-23pt} \\
\includegraphics[scale=0.35]{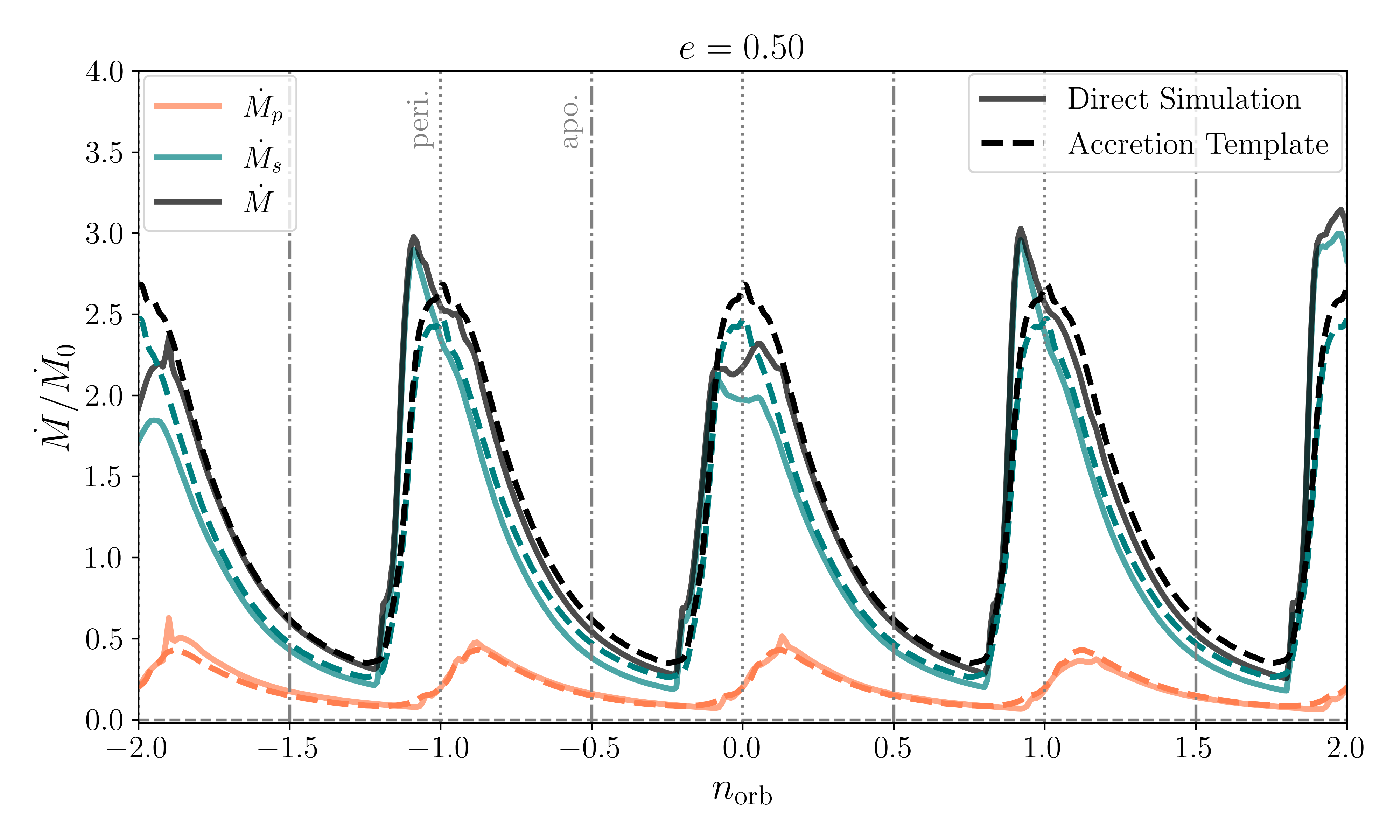} & 
\includegraphics[scale=0.35]{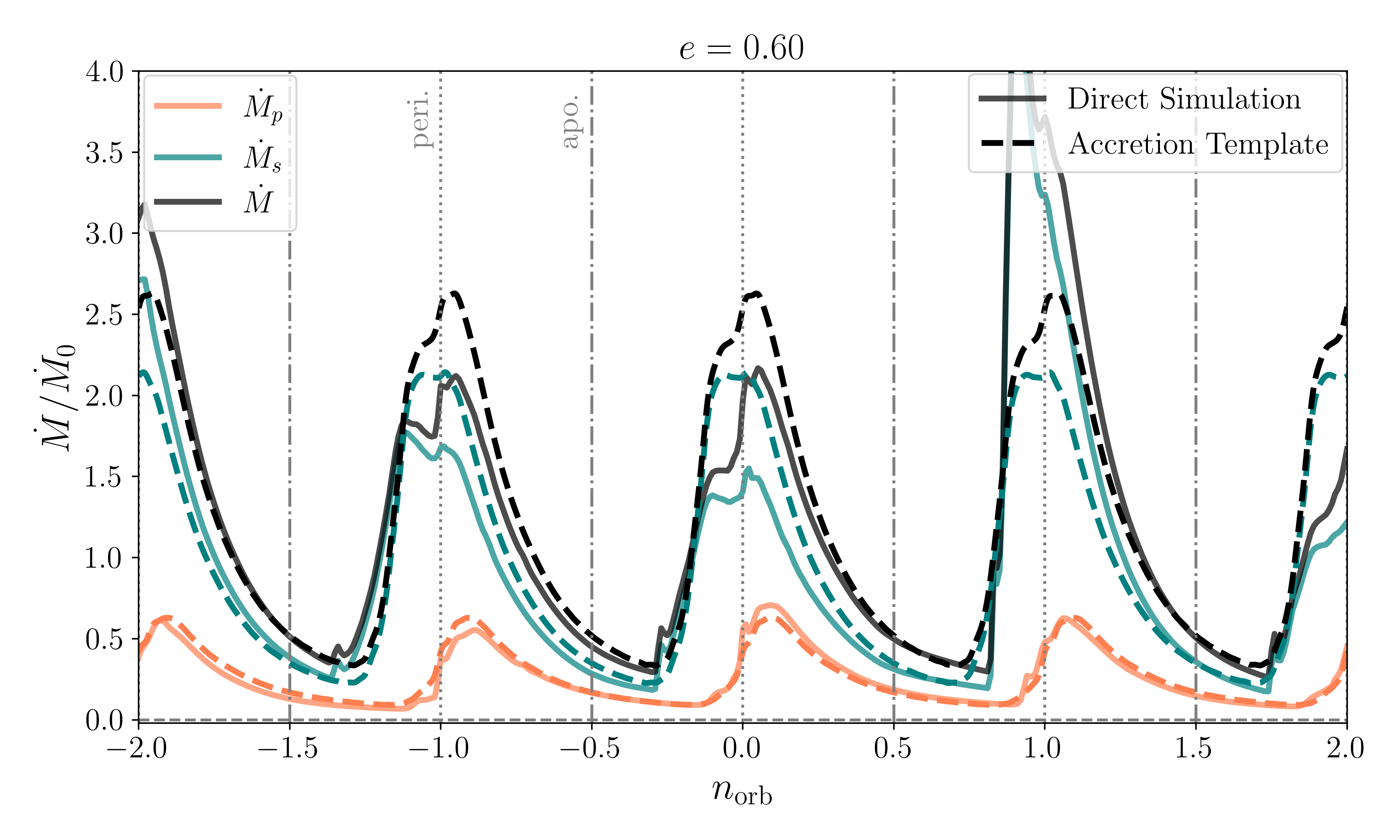} \\
\includegraphics[scale=0.35]{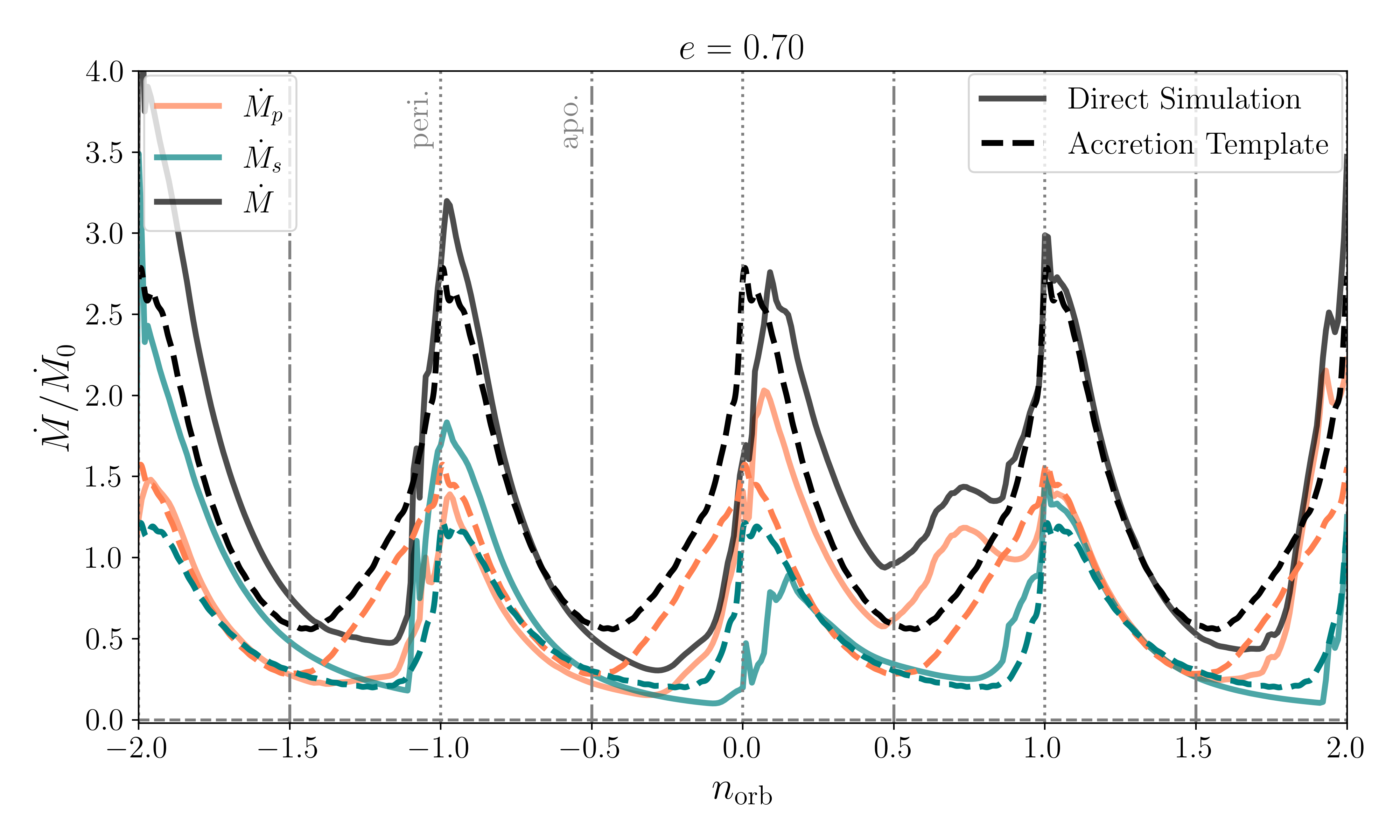} & 
\includegraphics[scale=0.35]{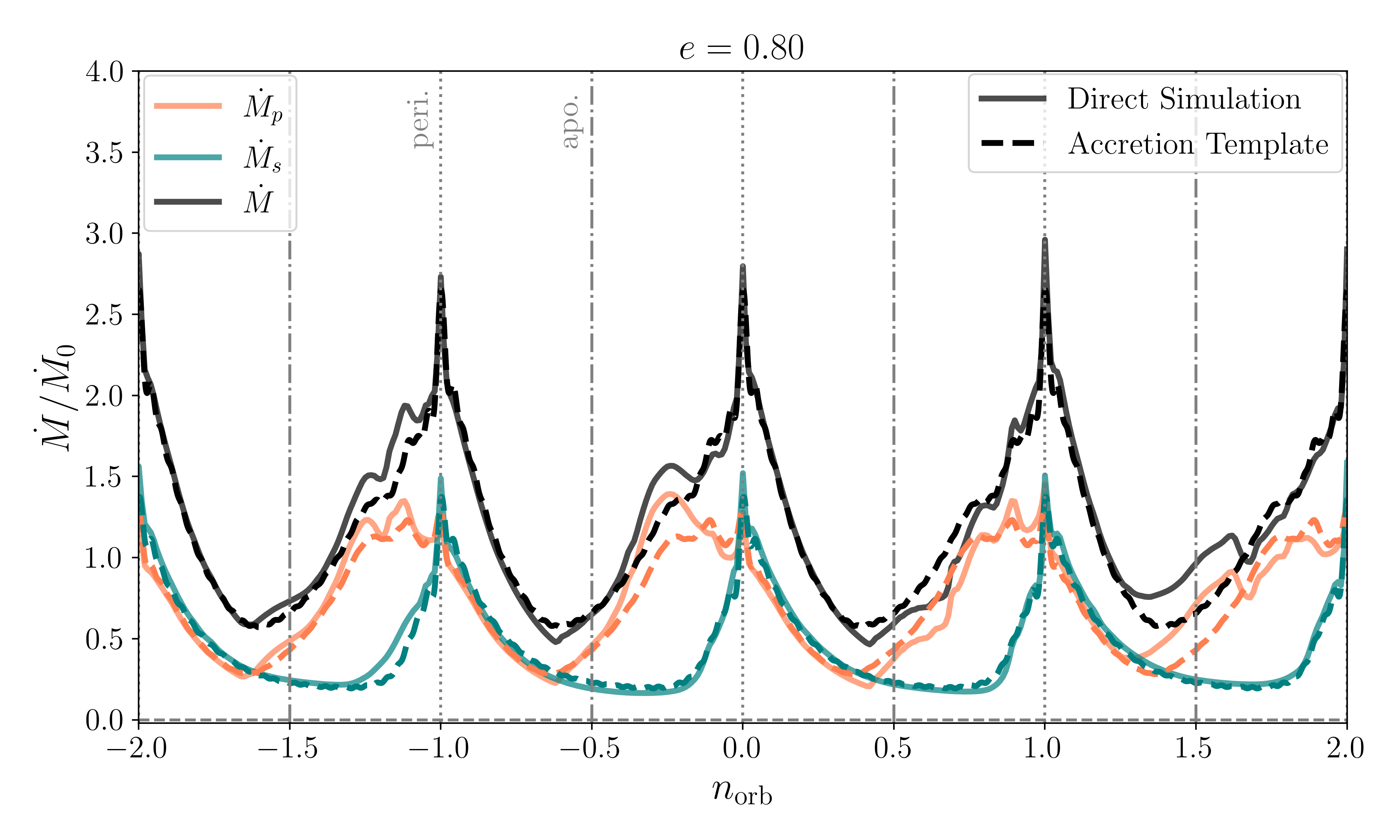}
% %
\end{array}$
\end{center}
\vspace{-15pt}
\caption{
Match of accretion-rate templates constructed with our code \texttt{binlite} (dashed curves), compared to the simulation output (solid curves). The total accretion rate onto the binary is plotted in black, while the breakdown of accretion onto each binary component is represented by the coral and teal curves. 
Accretion rates are plotted in units of the steady-state value $\dot{M}_0$.
The time series is displayed over four binary orbits comprising the center of a $\sim20$ orbit segment from which the Fourier reconstruction is built. Deviations between the reconstruction and simulation output for $e=0.5,0.6,0.7$ derive from inter-orbital variability (peak-to-peak variations) whose presence is apparent from the noisy region in the top-right portion of the corresponding periodogram in Figure \ref{Fig:2DPow}. The binary components have equal masses but the eccentric cavity causes preferential accretion that could equivalently be favoring either black hole. Hence, labeling of coral and teal curves can be interchanged.
}
\label{Fig:Mdot_prog_ALL}
\end{figure*}
%%%%%%%%%%%%%%%%%%%%%%%%%%%%%%%%%%%%%%%%%%%%%%%%

\section{Results}
\label{S:Results}

\subsection{Periodicity Analysis}
\label{S:Results_2DPow}

\paragraph{Prograde Periodogram}
The left panel of Figure \ref{Fig:2DPow} shows $|\mathcal{P}(e, \omega)|$ for prograde accretion computed via Eq. (\ref{eq:2Dpdgm}) over a grid of $600\times600$ values of $\omega$ and $e$. Bright colors denote significant power while dark colors denote lack of power at that corresponding point in parameter space -- the y-axis indicates the periodicity timescale in units of the binary orbital period, and the x-axis indicates binary eccentricity. 

For $e\lesssim 0.1$ we find that, in agreement with previous works \citep[\eg,][]{MunozLai_PulsedEccAcc:2016, MirandaLai+2017, Zrake+2021}, the power is concentrated at the orbital period and its harmonics, but is dominant in a small range of timescales centered around five binary orbital periods. This corresponds to the timescale for the cavity ``lump'' to circulate and periodically alter the feeding rate to the binary \citep{MacFadyen:2008, DHM:2013:MNRAS}. 

For $0.05 \lesssim e \lesssim 0.1$, power in the ``lump-timescale" splits into branches centered on the five-binary-orbit feature. This branching can be seen most easily in the higher frequency ($1/2.5 P^{-1}_b$) harmonic of the lump timescale on the left side of the left panel of Figure \ref{Fig:2DPow}. 
The dominant lump-branch drops in frequency from $1/5 P^{-1}_b$ to $1/4.5P^{-1}_b$ as $e$ increases from $0.05 \rightarrow 0.1$. At $e=0.1$ the lump branches vanish and are replaced by power that is almost entirely focused at the orbital timescale and its higher harmonics. Orbital period timescales then dominate for $e \gtrsim 0.1$.
At higher orbital eccentricities, particularly for $0.5 \lesssim e \lesssim 0.8$, the power is noisy at lower frequencies. This arises due to less stability in the accretion-rate time series from orbit to orbit. We demonstrate this further in the next section.

Our primary takeaway for the purpose of generating reconstructions of the accretion-rate time series is that the orbital period is a dominant feature for all eccentricities $e\geq0.1$, hence we choose $\Omega_b$ as our fundamental frequency for the Fourier reconstruction, Eqs. (\ref{Eq:F_rec}) and (\ref{Eq:F_amps}). For $e < 0.1$, we choose $\Omega_b/5$ as the fundamental frequency, but with more terms in the reconstruction.

\paragraph{Retrograde Periodogram}
The right panel of Figure \ref{Fig:2DPow} shows $|\mathcal{P}(e, \omega)|$ for retrograde accretion computed via Eq. (\ref{eq:2Dpdgm}) over a grid of $300\times300$ values of $\omega$ and $e$. A different version of this is also published in \citetalias{TiedeDOrazio:2023} (over a different range of timescales). Our main purpose for showing it here is to emphasize that the orbital timescale periodicity is strong for all eccentricities. However, in the retrograde case, a strong, two-times-orbital periodicity arises for $e\sim0.55$. Hence, for retrograde systems we choose the fundamental Fourier reconstruction frequency to be $\Omega_b/2$. Note that the retrograde periodogram is much less noisy than its prograde counterpart, indicating steadier accretion rate-times series, even for high eccentricities.

\subsection{Accretion-Rate Time Series}
\label{Ss:Results:Mdotcurves}

\paragraph{Prograde Time Series}
Figure \ref{Fig:Mdot_prog_ALL} presents example accretion-rate time series for prograde binaries with eight different values of orbital eccentricity. The solid lines show the accretion rates measured directly from the numerical calculations while the dashed lines are the accretion templates built from our Fourier reconstruction (Eqs. (\ref{Eq:F_rec}) and (\ref{Eq:F_amps})) using $\Omega = \Omega_b$ and a total of 30 Fourier components. Accretion rates onto each component are denoted in coral and teal, while the total is plotted in black. Vertical dotted lines denote the time of pericenter while vertical dot-dashed lines denote apocenter. Note that even though we have included enough Fourier components to capture sharp features in the time series (\eg, the bottom right panel of Figure \ref{Fig:Mdot_prog_ALL}), small deviations between the reconstructions (dashed) and the simulation (solid) are apparent for $e\gtrsim 0.5$. This is due to the inter-orbit variability which manifests as the noisy upper-right region in the prograde periodogram (left panel) of Figure \ref{Fig:2DPow}, \ie, at some eccentricities the accretion rate is less steady from one orbit to the next, affecting our reconstructions which are built from an average over $\sim 20$ orbits.

\begin{figure}
% \vspace{-10pt}
\begin{center}$
\begin{array}{c}
\hspace{-15pt}
\includegraphics[scale=0.35]{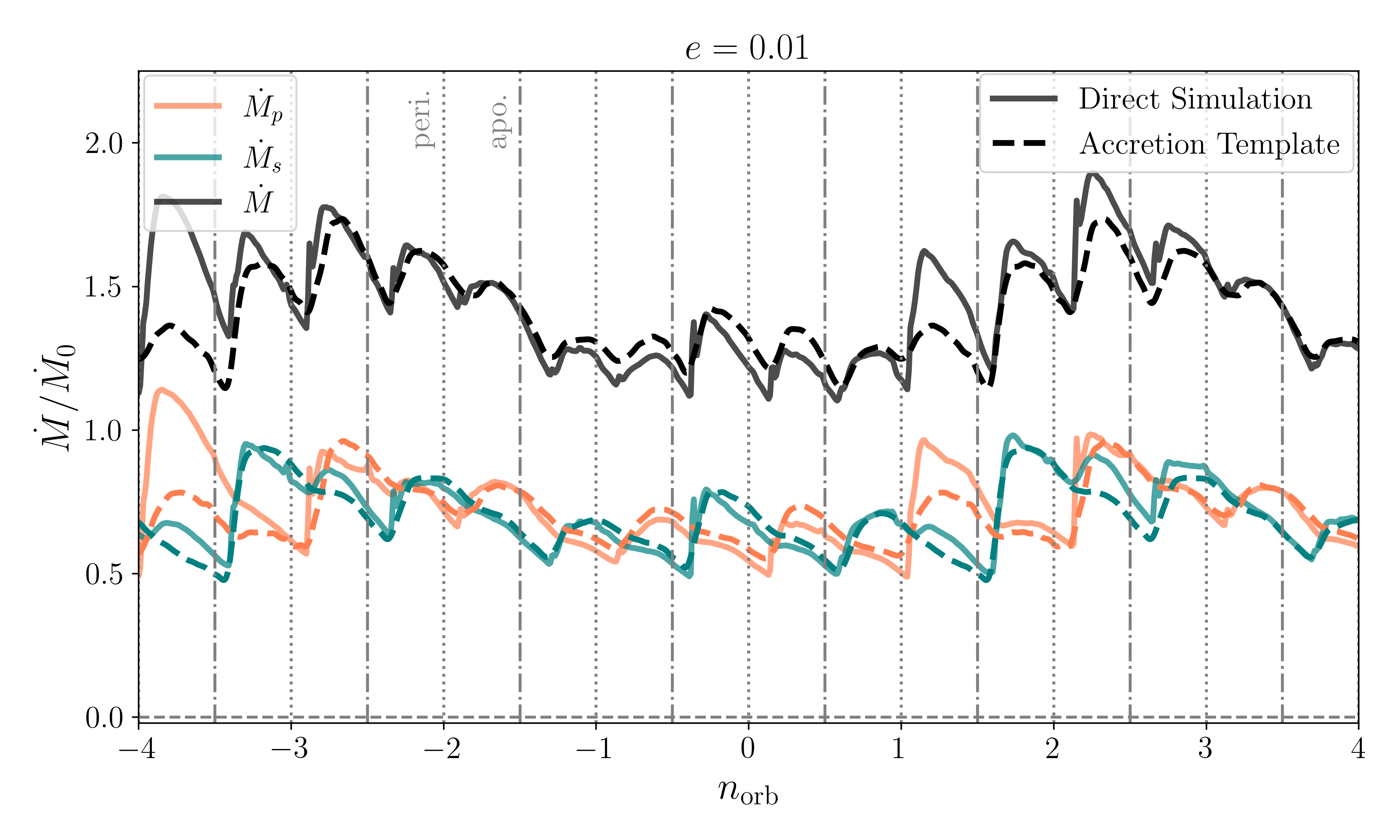}
% %
\end{array}$
\end{center}
\vspace{-15pt}
\caption{
The same as Figure \ref{Fig:Mdot_prog_ALL}, but for a prograde, $e=0.01$ system, where variability at $\Omega_b$ and $\Omega_b/5$ co-exist. 
}
\label{Fig:MdotRec_prog_e0p01}
\end{figure}

Figure \ref{Fig:MdotRec_prog_e0p01} demonstrates a reconstruction for a prograde binary with $e=0.01$, where the $\omega\sim\Omega_b/5$ periodicity of the cavity lump dominates. In this case the Fourier reconstruction uses $\Omega = \Omega_b/5$ and $60$ Fourier components. Here, the more complex nature of the variability is apparent in the less exact match of reconstruction and direct simulation (see again the more complex structure in the $e<0.01$ portion of the left periodogram in Figure \ref{Fig:2DPow}). Despite this, the reconstruction captures the main qualitative features of the time series, including crucially, the periodicity at both $\Omega_b$ and $\Omega_b/5$, and reliable reconstruction of the contribution of each component accretion rate to the total.

%%%%%%%%%%%%%%%%%%%%%%%%%%%%%%%%%%%%%%%%%%%%%%%%
%%% Prog-Lump Mdot Sim and Rec %%%
%%%%%%%%%%%%%%%%%%%%%%%%%%%%%%%%%%%%%%%%%%%%%%%%
%%%%%%%%%%%%%%%%%%%%%%%%%%%%%%%%%%%%%%%%%%%%%%%%

\begin{figure*}
\vspace{-10pt}
\begin{center}$
\begin{array}{cc}
\includegraphics[scale=0.35]{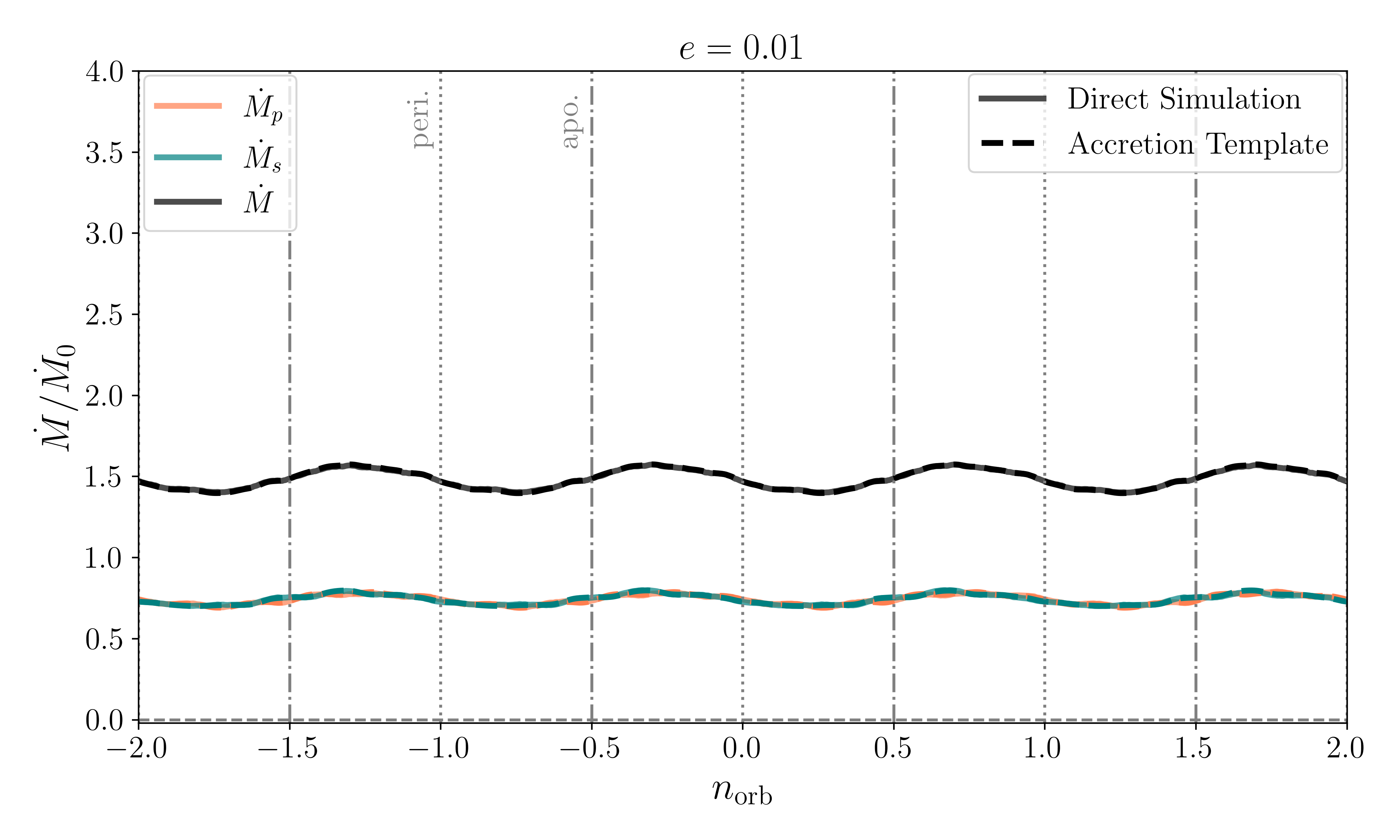} &
\includegraphics[scale=0.35]{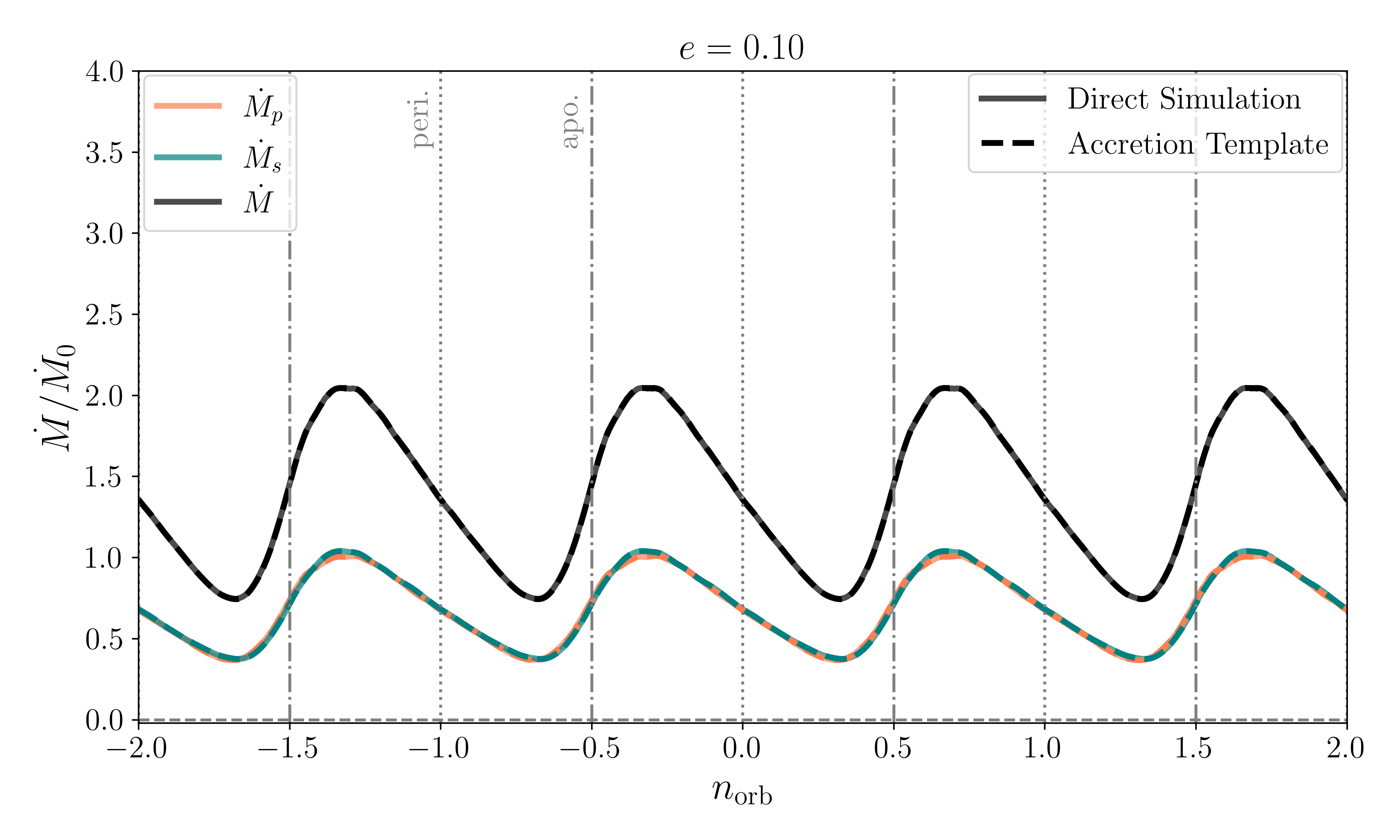} \\
\includegraphics[scale=0.35]{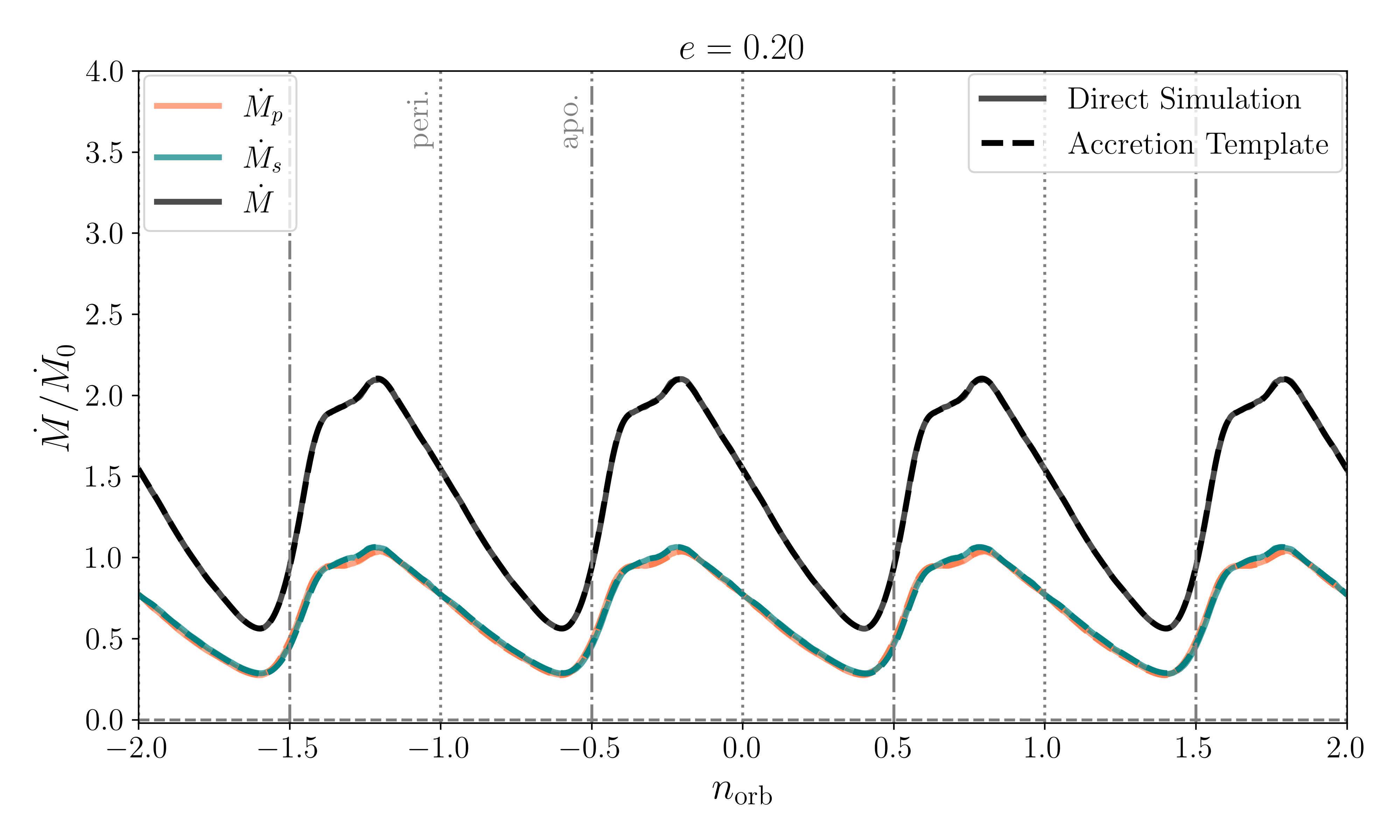} \vspace{-23pt} &
\includegraphics[scale=0.35]{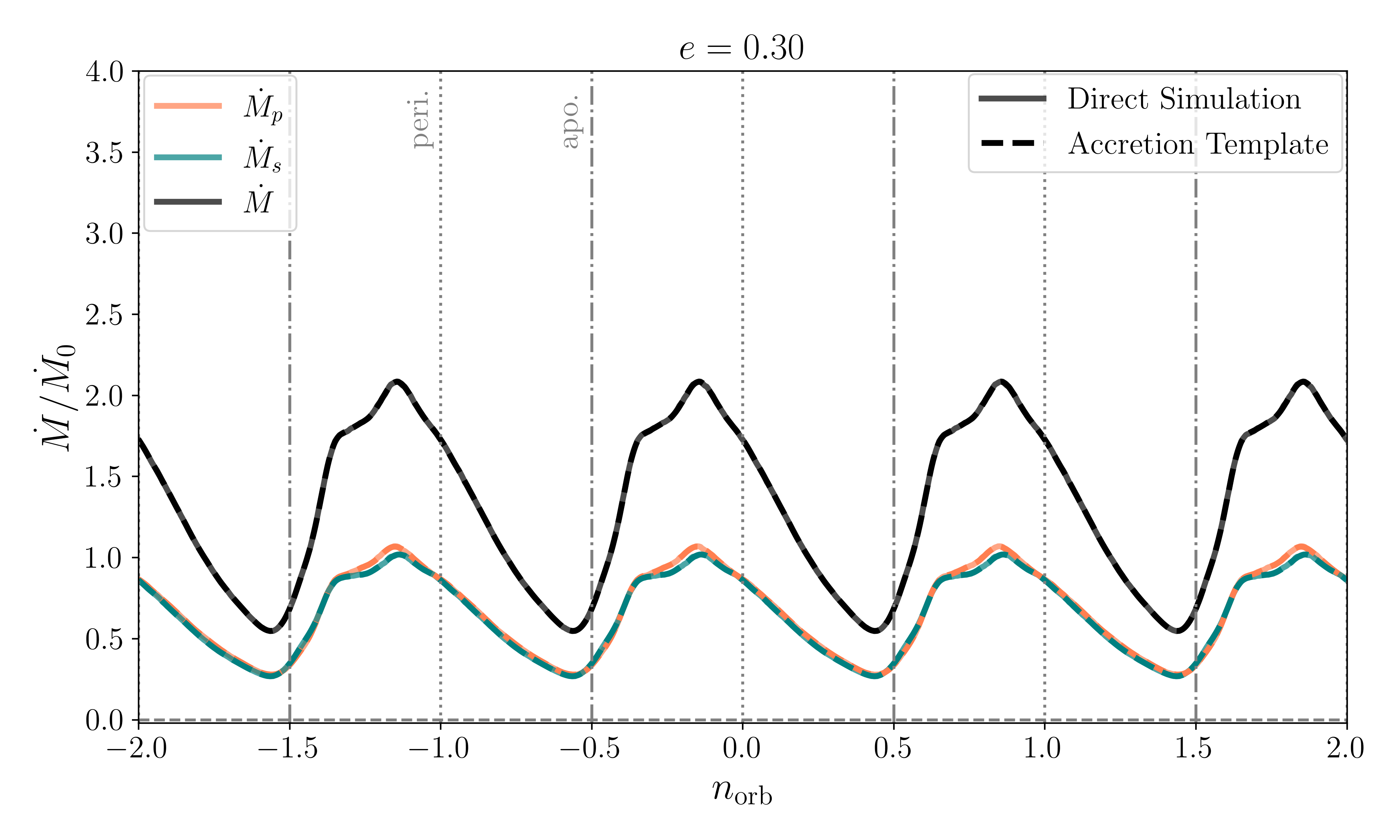} \\
\includegraphics[scale=0.35]{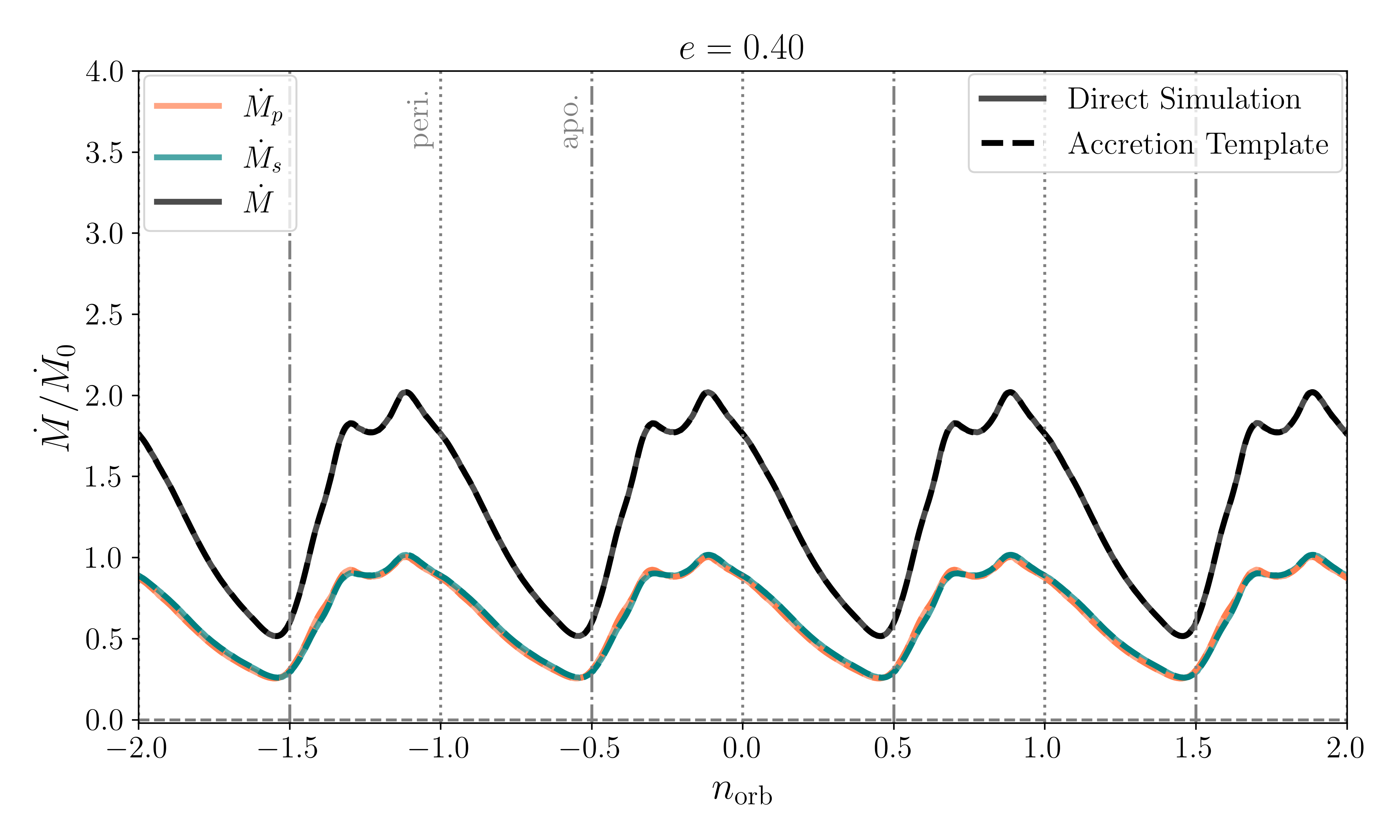} \vspace{-23pt} &
\includegraphics[scale=0.35]{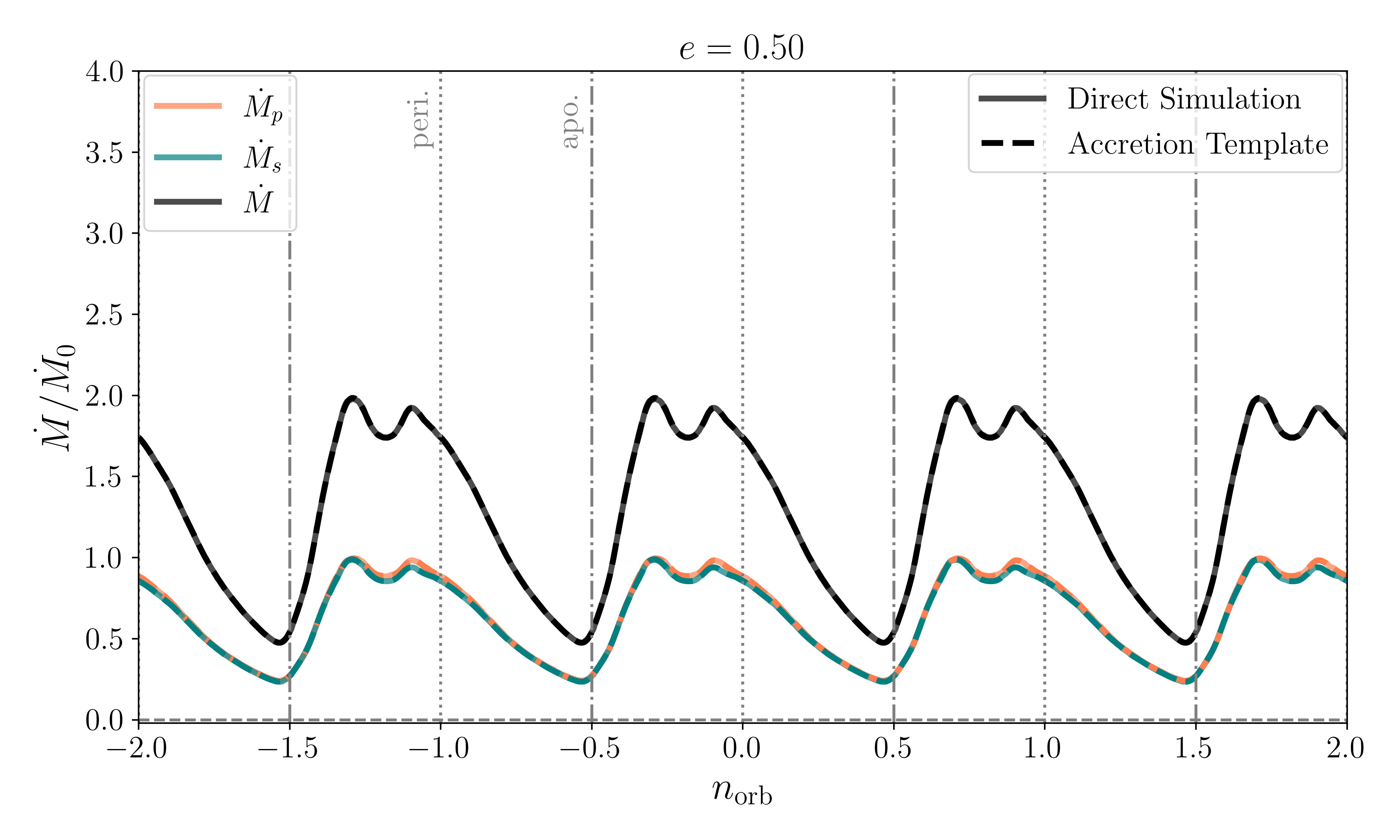} \\
\includegraphics[scale=0.35]{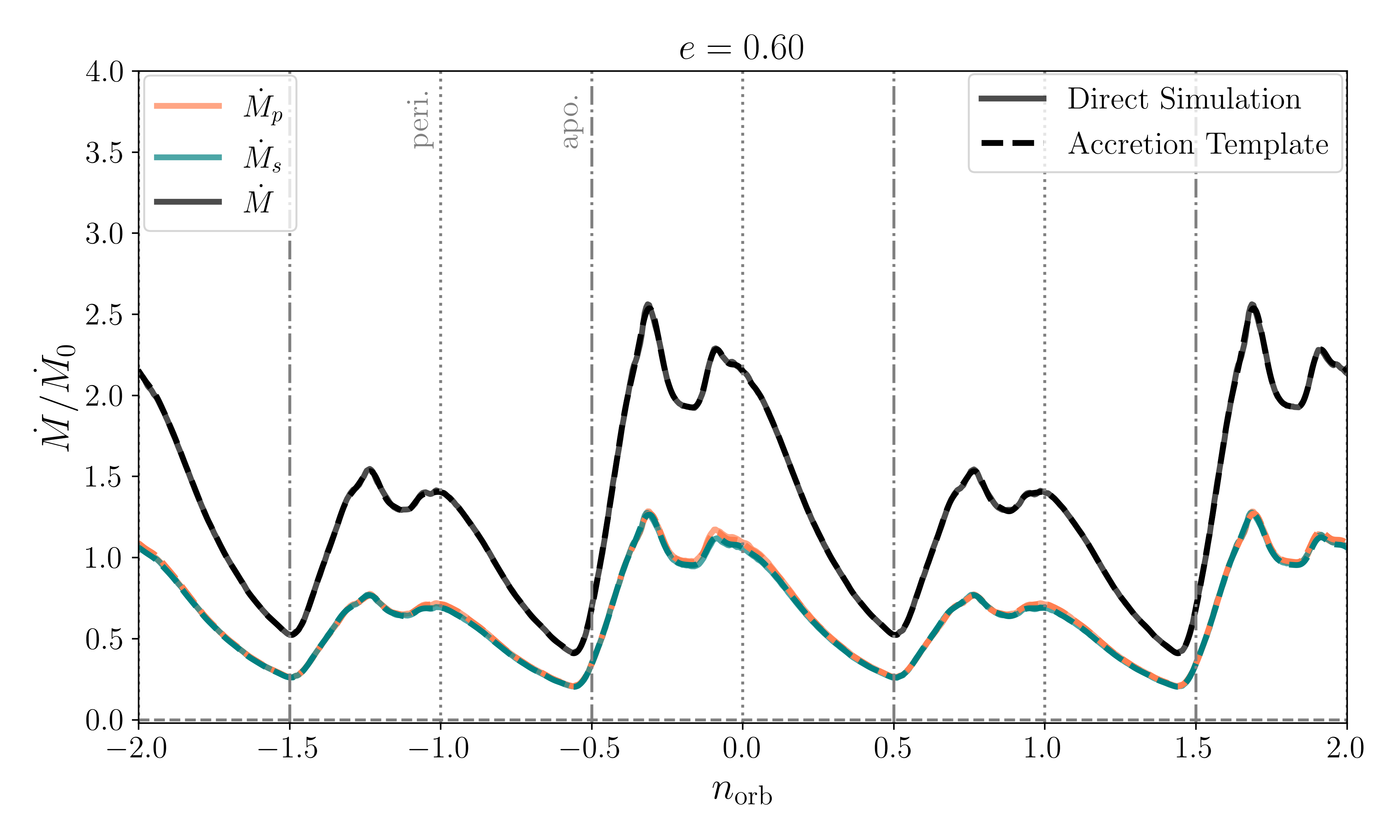} &
\includegraphics[scale=0.35]{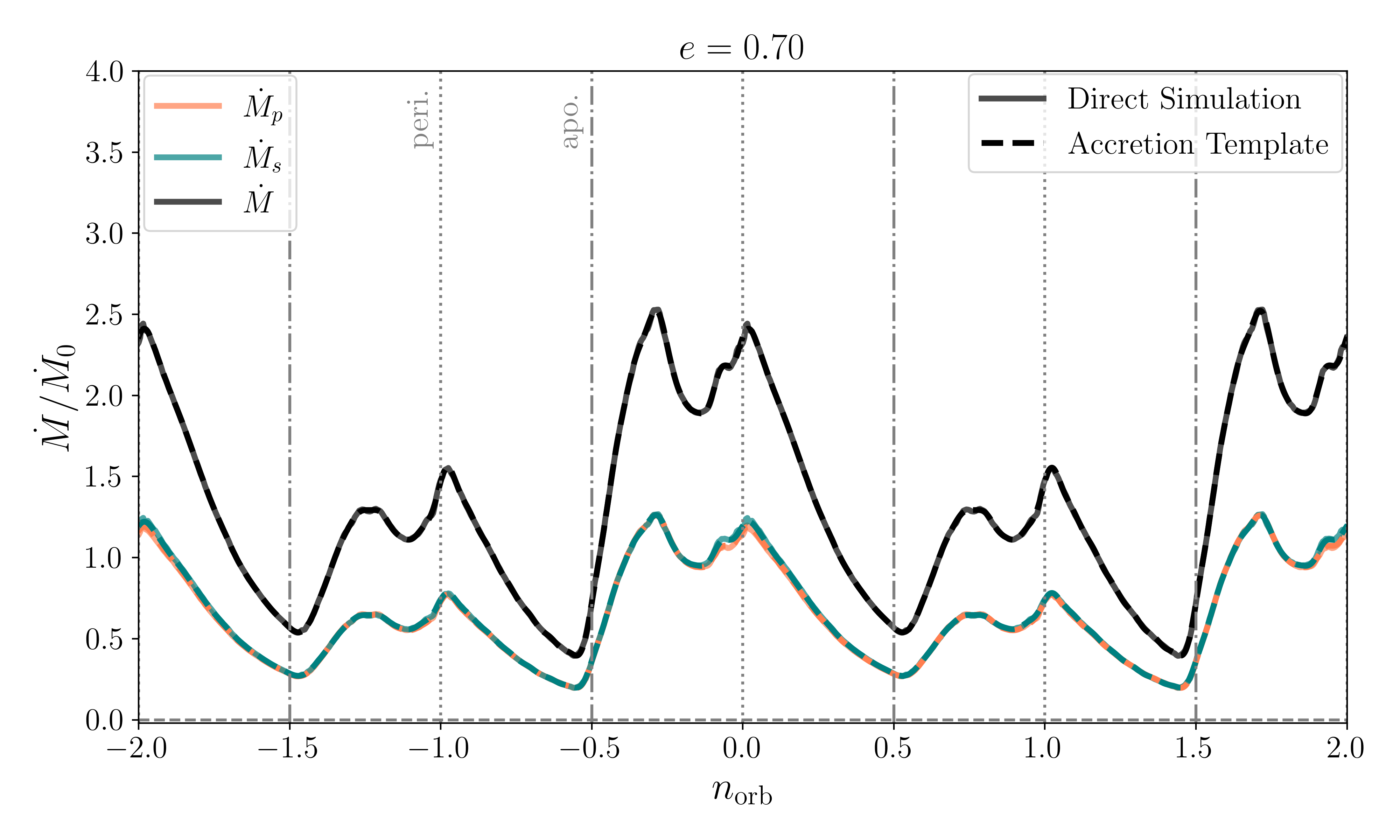} 
% %
\end{array}$
\end{center}
\vspace{-15pt}
\caption{
The same as Figure \ref{Fig:Mdot_prog_ALL} but for retrograde, eccentric binaries. Here the accretion rates onto the primary (coral) and secondary (teal) are nearly identical because there is no precessing, circumbinary cavity.
}
\label{Fig:MdotRec_Retro_Tot}
\end{figure*}

\paragraph{Retrograde Time Series}
Figure \ref{Fig:MdotRec_Retro_Tot} shows the reconstructed total, primary, and secondary accretion rates for retrograde binary-disk systems using $\Omega = \Omega_b/2$ and 30 terms in the Fourier reconstruction. The reconstructed and simulated cases capture both orbital and twice-orbital periodicity very well and result in nearly identical reconstructed vs. simulated curves. This is due to the much more steady nature of retrograde disk solutions across the parameter space. The total retrograde accretion rates are described further in \citetalias{TiedeDOrazio:2023}, while here we additionally show the component accretion rates. These are nearly identical to each other, as is expected for equal-mass binaries when no other asymmetry arises, \ie, the eccentric disk of the prograde case.

%%%%%%%%%%%%%%%%%%%%%%%%%%%%%%%%%%%%%%%%%%%%%%%%
%%% Retro Mdot Sim and Rec %%%
%%%%%%%%%%%%%%%%%%%%%%%%%%%%%%%%%%%%%%%%%%%%%%%%
%%%%%%%%%%%%%%%%%%%%%%%%%%%%%%%%%%%%%%%%%%%%%%%%

\subsection{Accretion-Rate Ratio $Q$}
\label{S:Qrat}

Over long enough timescales ($\mathcal{O}(10^2-10^3)$ binary orbits), the accretion rates onto equal-mass binaries will average to unity \citep[\eg,][]{Siwek:2023}\footnote{Though this may not be the case when binary eccentricity and disk eccentricity vectors are locked relative to each other, as could be the case for our $e\sim 0.2$ results -- this requires further investigation.}. However, when computing lightcurves, the relevant quantity is the accretion-rate ratio between the binary components over timescales spanning orbital periods. To quantify this, we define the ratio $Q = \left<\Mdot_2/\Mdot_1\right>$, averaged over an integer number of orbits for which the eccentric accretion-rate imbalance operates (much shorter than a disk precession timescale). 

\begin{figure}
% \vspace{-10pt}
\begin{center}$
\begin{array}{c}
\hspace{-10pt}
\includegraphics[scale=0.5]{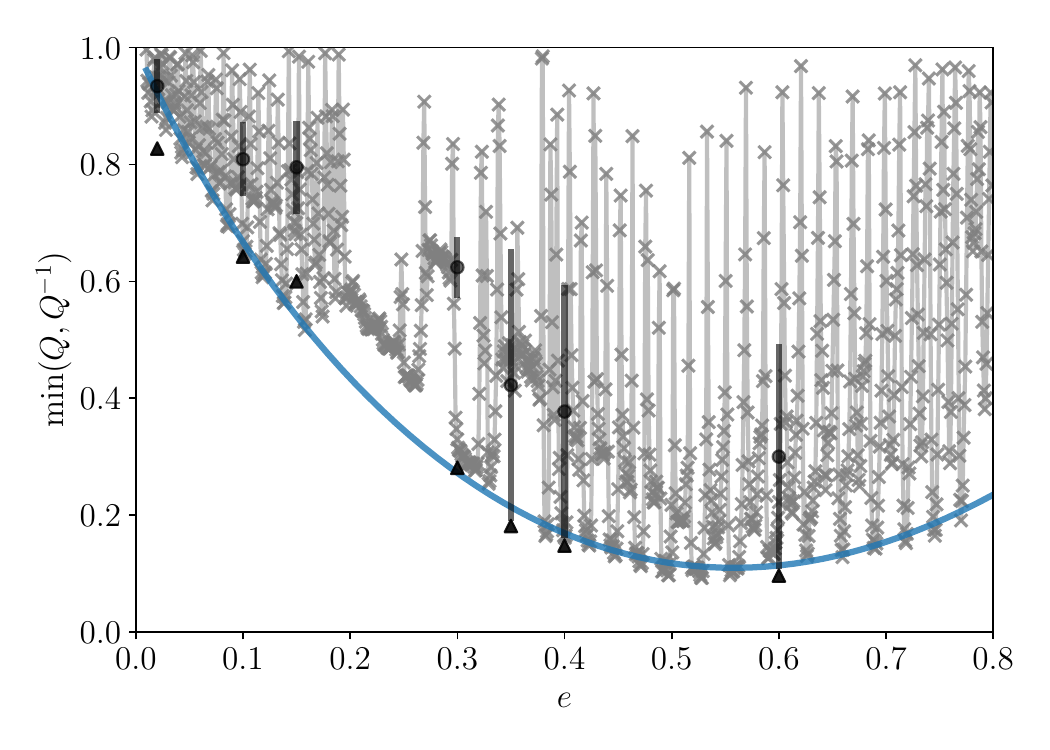} 
% %
\end{array}$
\end{center}
\vspace{-15pt}
\caption{
The ratio $Q \equiv \left<\Mdot_2/\Mdot_1\right>$ of accretion rates onto each binary component, averaged over a duration shorter than the disk precession frequency. The grey x's are measured from our reconstruction while the blue line is an analytic approximation for the extreme values. As a check of the values derived from the eccentricity sweep, the black points show results measured from simulations with fixed eccentricities -- black points with error bars represent the average and standard deviation of $\min\left(Q, Q^{-1} \right)$ over a precession period of the disk , and the triangles denote the smallest values over the same range.
}
% $\min\left(Q, Q^{-1} \right)$ is the average of $\left<Q\right>$ over one half of the disk precession period and \left<Q^{-1}\right> over the other half -- so it does not go to unity.
\label{Fig:Qrat}
\end{figure}

The ratio of component accretion rates, $Q$ is always unity for retrograde, equal-mass binaries (see Figure \ref{Fig:MdotRec_Retro_Tot}) but is a function of eccentricity for prograde binaries. In Figure \ref{Fig:Qrat} we present $Q$ measured from the prograde simulations as a function of binary eccentricity. The grey x's represent 1000 values drawn from the reconstructed accretion rates averaged over 5 orbits, sampled evenly in $e$. Because preferential accretion trades between primary and secondary, we present the minimum of $Q$ and $Q^{-1}$, with both values being valid choices when modelling equal-mass binaries considered here. 
When plotted this way, the extreme values of the ratio of accretion rates can be approximated by a simple function of eccentricity, inspired by the ratio of pericenter and apocenter distances of the binary components,
\begin{eqnarray}
\label{Eq:Qfit}
        Q_{\min} &\approx& \frac{1-\mathcal{P}(e)e}{1+\mathcal{A}(e)e} \\
        \mathcal{P}(e) &=& 2 - e^2 - 2e^3  \nonumber\\ %\quad
        \mathcal{A}(e) &=& 2 + e^2, \nonumber 
\end{eqnarray}
which is drawn as the solid blue line in Figure \ref{Fig:Qrat}. 

For most binary eccentricities the accretion-rate periodically switches back and forth between favoring each of the binary components. Hence, the ratio of accretion rates used to compute lightcurves can take values between $Q_{\min}$ and $Q^{-1}_{\min}$. As can be seen from the density of grey x's in Figure \ref{Fig:Qrat}, the binary spends more time accreting at some ratios than others: $Q=1$ is sparsely sampled because this value is encountered during the relatively rapid stage where the accretion rate switches from favoring one component to favoring the other. 
The black points and associated error bars show the average accretion ratio from a set of constant-eccentricity verification runs (detailed in \citetalias{DOrazioDuffell:2021}) and the black triangles illustrate the smallest value over the run. Those values of $Q$ that lie within the error bars from these constant-eccentricity runs provide a good indicator for where the binary spends most of its time accreting.

For intermediate values of eccentricity, the disk is more symmetric around the origin and is either slowly precessing or not precessing at all. This results in regions with narrow spreads of accretion-rate ratio $Q$, seen as the clustering of grey x's between $0.018 \lesssim e \lesssim 0.38$ in Figure \ref{Fig:Qrat}. While the spread of $Q$ values is much smaller in this range than outside of it, $Q$ is not continuous in $e$ here and jumps through four different states of monotonically varying $Q(e)$, while also experiencing islands of oscillating disk solutions near the transitions between these four states. This behavior is likely due to locking of the angle between disk and binary eccentricity and would be worth studying further for elucidating the binary+disk dynamics in this regime and its relation to observability via, \eg, accretion variability.

%%%%%%%%%%%%%%%%%%%%%%%%%%%%%%%%%%%%%%%%%%%%%%%%
%%% Q(e) %%%
%%%%%%%%%%%%%%%%%%%%%%%%%%%%%%%%%%%%%%%%%%%%%%%%
%%%%%%%%%%%%%%%%%%%%%%%%%%%%%%%%%%%%%%%%%%%%%%%%

\section{Application: Construction of Boosted and Lensed Lightcurves}
\label{S:Application}

As an example-use case and to demonstrate the wide range of lightcurve shapes that can arise from accreting, eccentric binaries, we develop a simple procedure for converting the accretion-rate time series of the previous section to a rest-frame flux. Primarily for application to accreting black hole binaries, we then convert this to an observer-dependent flux by including relativistic orbital Doppler boosting of emission emanating from the minidisks around each binary component \citep{PG1302Nature:2015b}, as well as binary self-lensing in the point mass, point-source limit \citep{DoDi:2018, Spikey:2020}.

Doppler boosting and binary lensing have been put forth as mechanisms for causing unique periodic variability in accreting SMBHB systems \citep{PG1302Nature:2015b, DoDi:2018, Spikey:2020, JordylensI:2022, IngramLens+2021, DOrazioCharisi:2023, KrauthLens+2023}. Modeling these signatures consistently requires knowledge of the fraction of light coming from each of the minidisks and the CBD. It should also be combined with the intrinsic variability of the source in order to understand systems where both hydrodynamical variability as well as observer dependent effects are jointly operating.\footnote{A combination of Doppler boosting and hydrodynamic variability near merger is simulated in \citet{Yike:2018}.} Hence, the reconstructed accretion-rate time series of the previous section allow us to significantly build upon toy models for the variability of accreting SMBHBs.

We model the emission from the accreting binary in a given frequency band as a constant specific flux from the circumbinary disk, $F^0_{\nu,\mathrm{CBD}}$, plus time-dependent emission from the primary and secondary minidisks. We model the minidisk emission as a constant, average flux times a time-dependent function $F^0_{\nu,1} p(t)$ and $F^0_{\nu,2} s(t)$, where $p(t) \equiv \dot{M}_1/\left<\dot{M}_1+\dot{M}_2\right>$ and $s(t)\equiv \dot{M}_2/\left<\dot{M}_1+\dot{M}_2\right>$ are the reconstructed time-variable accretion rates computed with \codename~(Section  \ref{Ss:Results:Mdotcurves}) and normalized by the average total accretion rate at that eccentricity.

\paragraph{Lightcurve Generation}
With boosting and lensing taken into account the total observed flux is
\begin{eqnarray}
    F_{\nu} = F^0_{\nu,1} \mathcal{D}_1 \Mag_1 p(t) + F^0_{\nu,2}  \mathcal{D}_2 \Mag_2 s(t) + F^0_{\nu,\mathrm{CBD}},
\end{eqnarray}
where $\mathcal{D}_i \equiv D^{3-\alpha}_i$ is the time-dependent Doppler-boost magnification for Doppler factor $D_i$ and frequency-dependent log-spectral slope $\alpha$, in the observing band (assumed here to be the same for all disk components), while $\Mag_i$ is the time-dependent lensing magnification for the specified binary component \citep[see][]{DoDi:2018, Spikey:2020}. 
Defining 
\begin{eqnarray}
    F^0_{\nu,\mathrm{Tot}} &\equiv& F^0_{\nu,1} + F^0_{\nu,2} + F^0_{\nu,\mathrm{CBD}}, \nonumber \\
    \chi_{\nu,1} &\equiv& F^0_{\nu,1}/F^0_{\nu,\mathrm{Tot}}, \nonumber \\
    \chi_{\nu,2} &\equiv& F^0_{\nu,2}/F^0_{\nu,\mathrm{Tot}},
\end{eqnarray}
we can write the observed in-band flux, normalized to the total average (rest frame) flux as,
\clearpage
\begin{eqnarray}
    \frac{F_{\nu}}{\left<F^0_{\nu,\mathrm{Tot}}\right>} &=& \chi_{\nu,1}  \mathcal{D}_1 \Mag_1 p(t) + \chi_{\nu,2} \mathcal{D}_2 \Mag_2 s(t) \nonumber \\   
    &+& (1-\chi_{\nu,1}-\chi_{\nu,2}), 
    \label{Eq:Magnfy}
\end{eqnarray}
which reduces to
\begin{eqnarray}
    \frac{F_{\nu}}{\left<F^0_{\nu,\mathrm{Tot}}\right>} = (1-\chi_{\nu,2})  \mathcal{D}_1 \Mag_1 p(t) + \chi_{\nu,2}  \mathcal{D}_2 \Mag_2 s(t),
\end{eqnarray}
for a simplified case where the minidisks are assumed to outshine the CBD.

Hence, we need to know the relative fluxes, in a specified frequency band, from each minidisk and the circumbinary disk. This requires knowing the disk spectra, which depends on physics of radiative energy balance in the accretion flow not captured by our isothermal simulations. For simplicity, we treat the circumbinary disk and both minidisks as separate components and approximate the spectra of each component with composite black-body spectra of the optically thick alpha-disk solutions. In this case, the spectrum is set by each disk temperature profile, $T(r) \propto (M\Mdot/r^3)^{1/4}$, which, for equal-mass binary components differs between minidisks only through the accretion rate, and by an extra mass factor $2^{1/4}$ for the circumbinary disk surrounding the total binary mass.

This allows us to compute spectra of each of the disk components by choosing a total accretion rate through the CBD, a binary mass ratio, and using the split in accretion rates onto the binary components measured from the reconstructed accretion rates of Section \ref{S:Results}. Over long enough timescales, the average accretion rate through the circumbinary disk is split evenly for equal mass binary components. However, as discussed in Sections \ref{Ss:Results:Mdotcurves} and \ref{S:Qrat}, for prograde eccentric binaries, this balance can be shifted back and forth between components over periods of $\mathcal{O}(100)$ binary orbits as the eccentric disk precesses with respect to the binary argument of pericenter. 
To quantify this, we use the accretion-rate ratio explored in Section \ref{S:Qrat} and Figure \ref{Fig:Qrat}, $Q \equiv \left<\Mdot_2/\Mdot_1\right>$. We require that the minidisks are fed by a circumbinary disk with total mass-accretion rate $\Mdot_{\rm CBD}$. Then the measured $Q$ and choice of $\Mdot_{\rm CBD}$ specify the system:
\begin{eqnarray}
    \Mdot_1 &=& \Mdot_{\rm CBD} \left( 1 + Q\right)^{-1}; \quad \Mdot_2 = Q\Mdot_1 ,
\end{eqnarray}
so that, evaluated at the same radius,
\begin{eqnarray}
    T_{\rm CBD} = 2^{1/4}(1+Q)^{1/4}T_{1}; \quad T_2 = Q^{1/4}T_1,
\end{eqnarray}
where we have assumed that the binary components have equal masses in the second line. The average fluxes from each disk component are,
\begin{widetext}
\begin{eqnarray}
    \label{Eq:Fnus}
    F^0_{\nu,1} &=& \frac{2 \pi \cos{I}}{d^2} \int^{r_{o,1}}_{r_{i,1}} B_{\nu}\left[  T_{i,1}\left(\frac{r}{r_{i,1}}\right)^{-3/4} \right] \ r dr  \\   \nonumber %\label{Eq:Fnus}
    F^0_{\nu,2} &=& \frac{2 \pi \cos{I}}{d^2} \int^{r_{o,2}}_{r_{i,2}} B_{\nu}\left[ Q^{1/4} T_{i,1}\left(\frac{r}{r_{i,1}}\right)^{-3/4} \right] \ r dr     \\
    F^0_{\nu,\mathrm{CBD}} &=& \frac{2 \pi \cos{I}}{d^2} \int^{r_{o,\mathrm{CBD}}}_{r_{i,\mathrm{CBD}}} B_{\nu}\left[2^{1/4} (1+Q)^{1/4} T_{i,1}\left(\frac{r}{r_{i,1}}\right)^{-3/4} \right] \ r dr, \nonumber
\end{eqnarray}
\end{widetext}
for a source at distance $d$ and a common disk-inclination angle $I$. For the examples here, we assume that the CBD extends from $r_{i,\mathrm{CBD}}=2a$ to $r_{o,\mathrm{CBD}}=100a$ and that the minidisks extend from the Schwarzschild inner-most stable circular orbit (ISCO) of the black hole, \eg, $r_{i,1}=6GM_1/c^2$, to the tidal truncation radius, $r_0 \approx 0.27 a = 0.27(\Omega_b)^{-2/3}(GM)^{1/3}$ \citep{Roedig+Krolik+Miller2014}. The quantity $T_{i,1}$ is the temperature in the primary minidisk at $r_{i,1}$. We emphasise that in this model, Eqs. (\ref{Eq:Fnus}) set the average flux scale for each disk component while time variability comes from the \codename~accretion-rate time series of Section \ref{Ss:Results:Mdotcurves}.

\begin{figure*}
% \vspace{-10pt}
\begin{center}$
\begin{array}{cc}
\hspace{-10pt}
\includegraphics[scale=0.35]{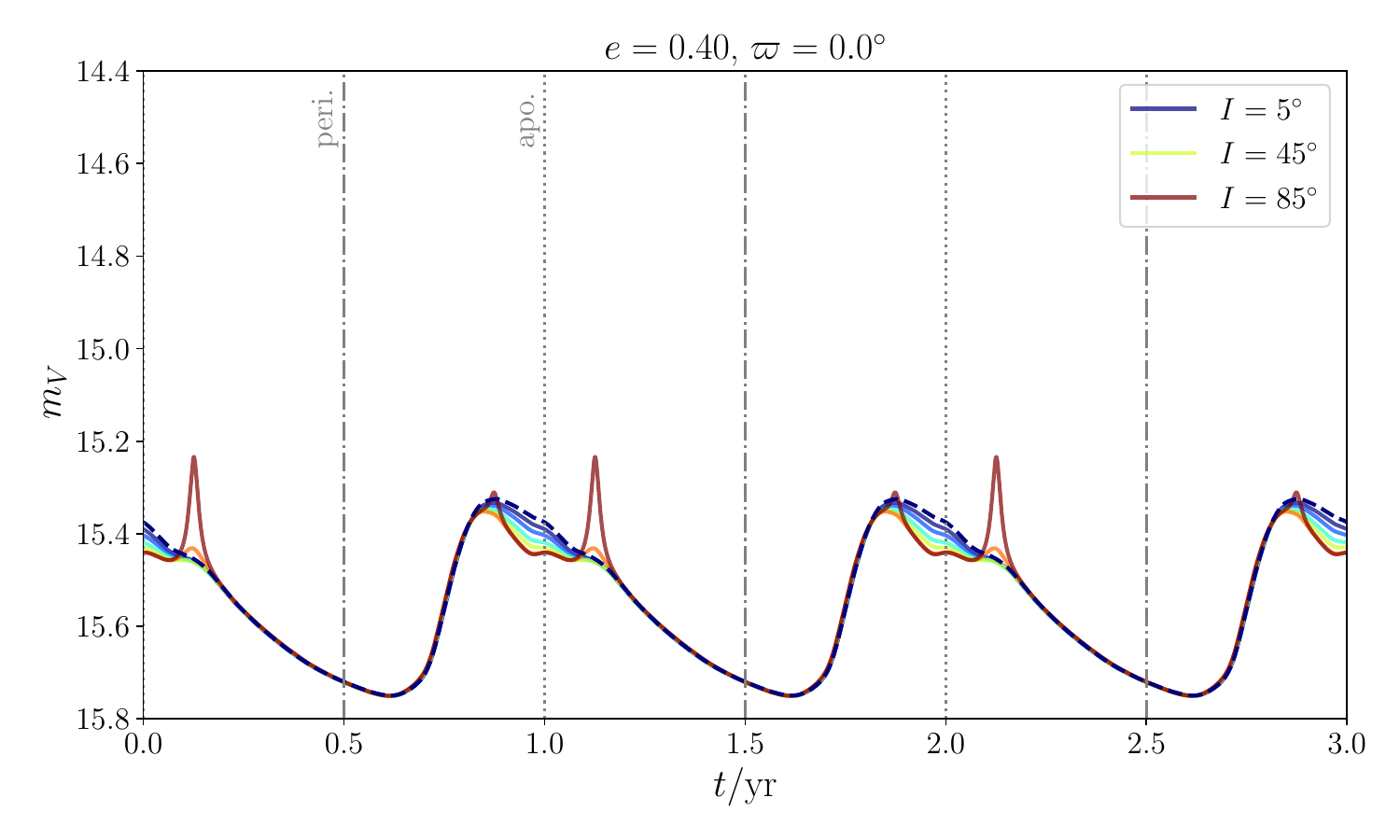} &
\includegraphics[scale=0.35]{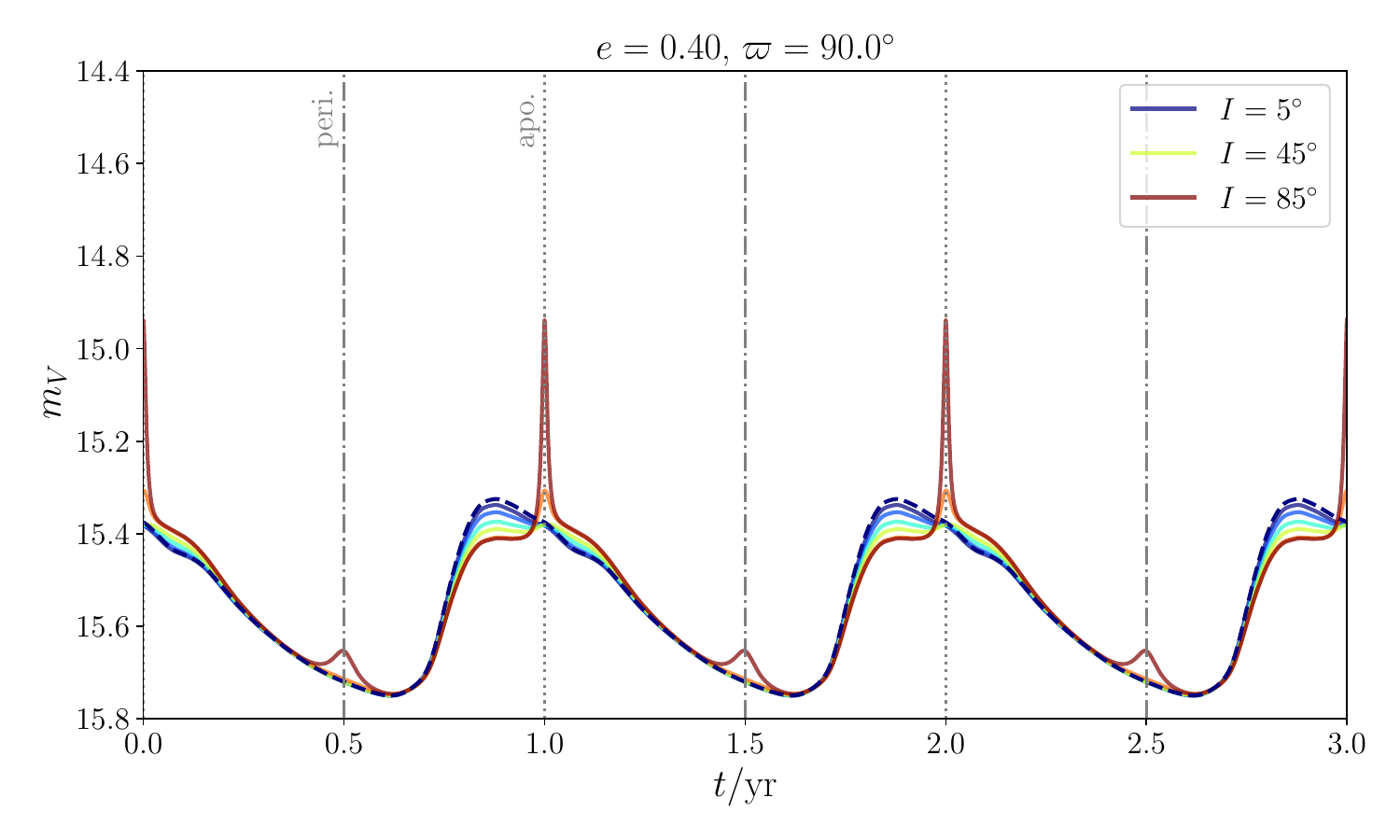} \\
\hspace{-10pt}
\includegraphics[scale=0.35]{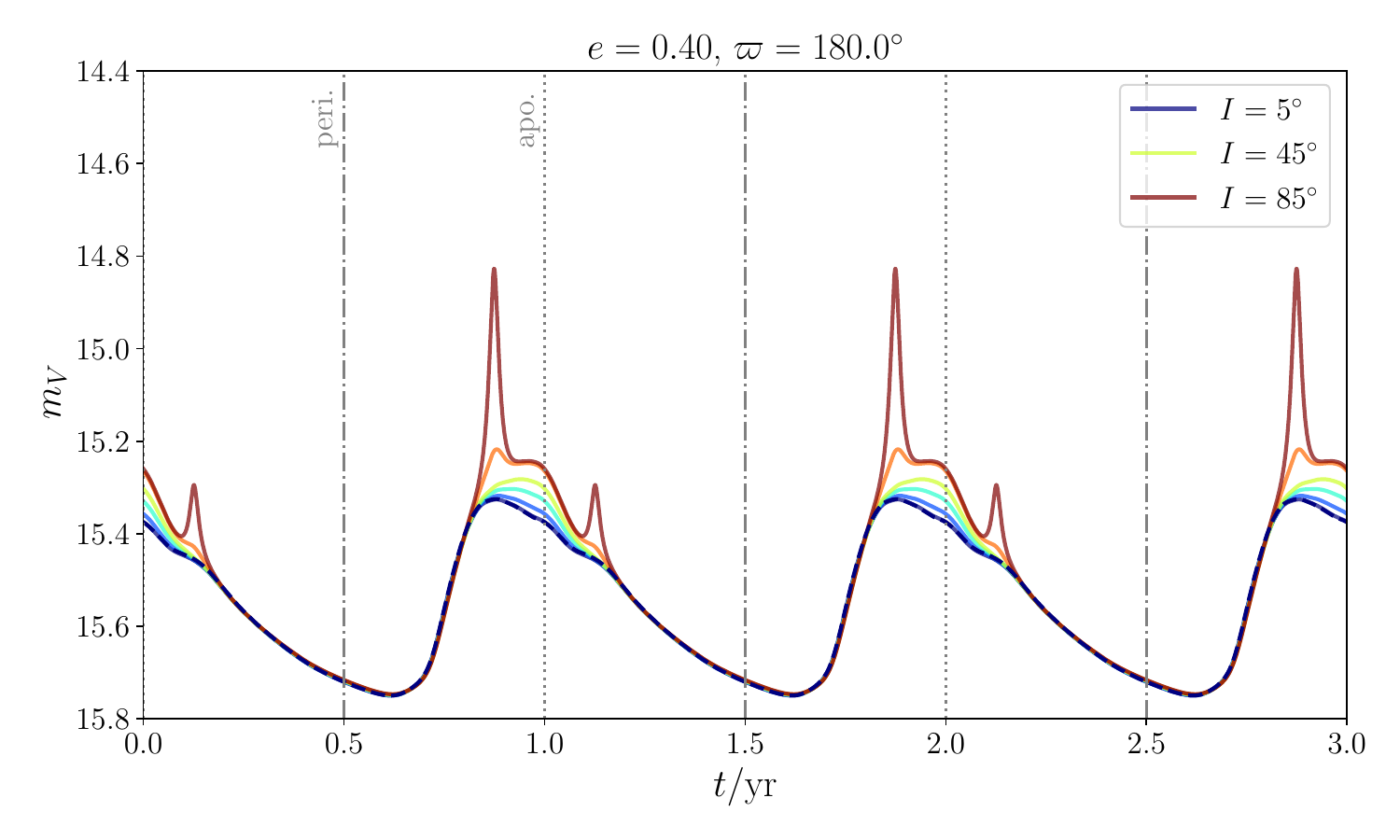} &
\includegraphics[scale=0.35]{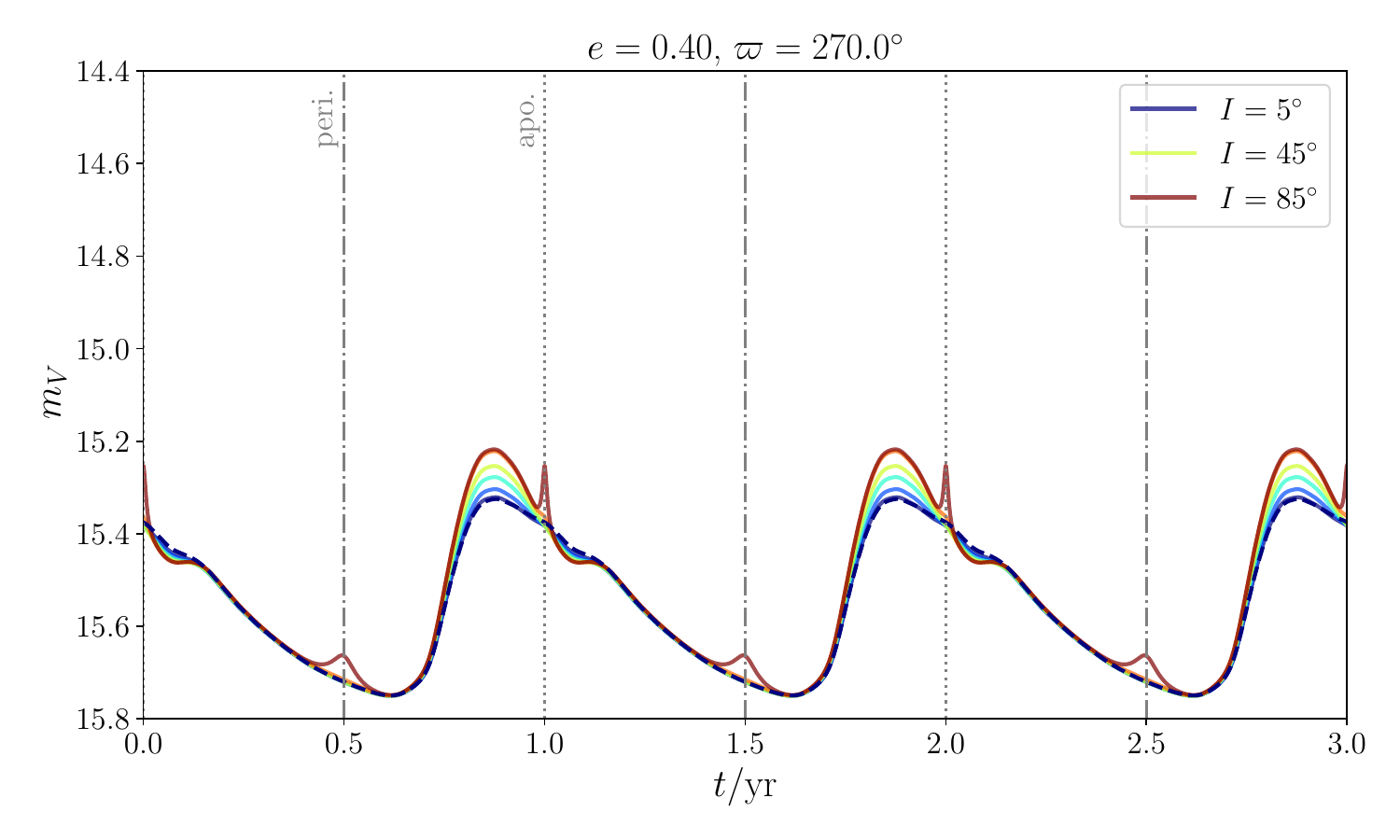} 
\end{array}$
\end{center}
\vspace{-15pt}
\caption{
V-band apparent magnitude lightcurves for prograde systems, (see Section \ref{S:Application}). Each panel is drawn for a different azimuthal viewing angle of the binary orbit relative to the argument of pericenter, $\varpi$, and each differently colored line is for a different binary inclination to the line of sight. The dashed line represents a face-on binary and so exhibits purely accretion-rate-induced flux variability (compare to Fig.~\ref{Fig:Mdot_prog_ALL}). The strongest Doppler and lensing effects arise for the red line, drawn for a binary inclined close to the line of sight, $I=85^{\circ}$. Each panel assumes a binary with $M=2 \times 10^9\Msun$ and $P=1$yr, at a distance of $1.5$~Gpc, and accreting at $10\%$ of the Eddington rate.
}
\label{Fig:LCs_pom}
\end{figure*}

\begin{figure*}
% \vspace{-10pt}
\begin{center}$
\begin{array}{cc}
\hspace{-10pt}
\includegraphics[scale=0.35]{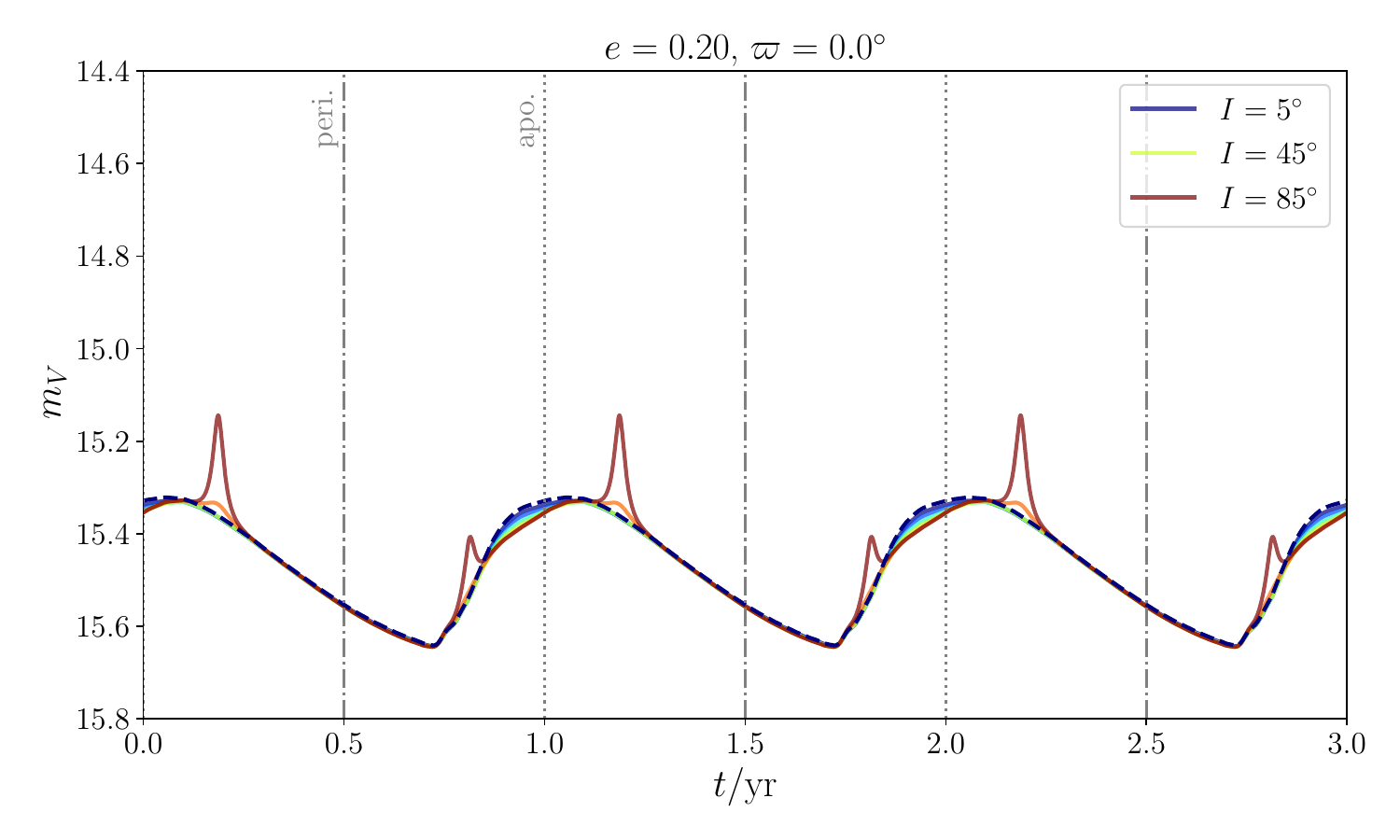} &
\includegraphics[scale=0.35]{fin_fig/LCmagV_vs_t_e0.40_omega0.00_chi10.307801_chi20.160371_Mbin9.30103_Pbin1yr_fEddCBD0.1_t00.00P} \\
\hspace{-10pt}
\includegraphics[scale=0.35]{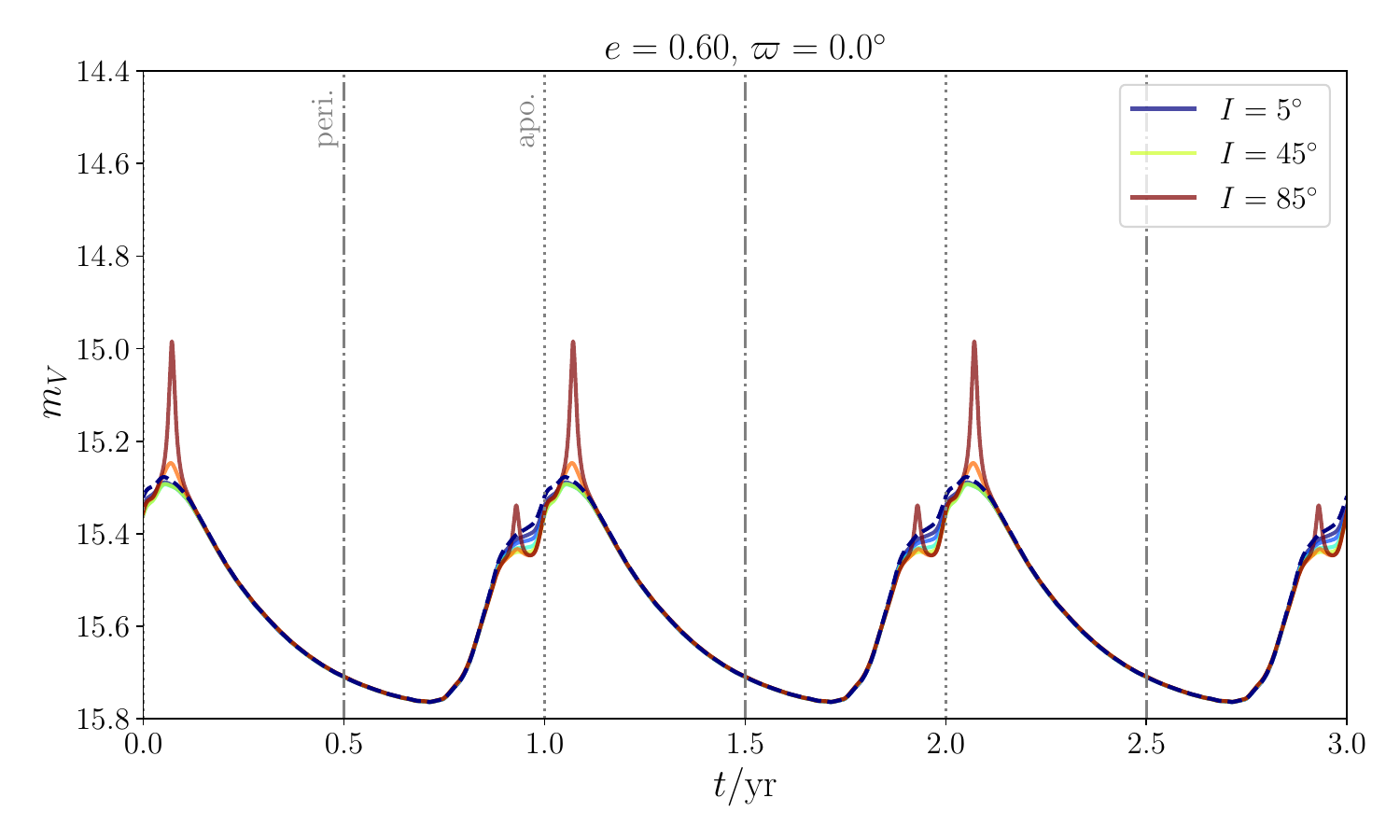} &
\includegraphics[scale=0.35]{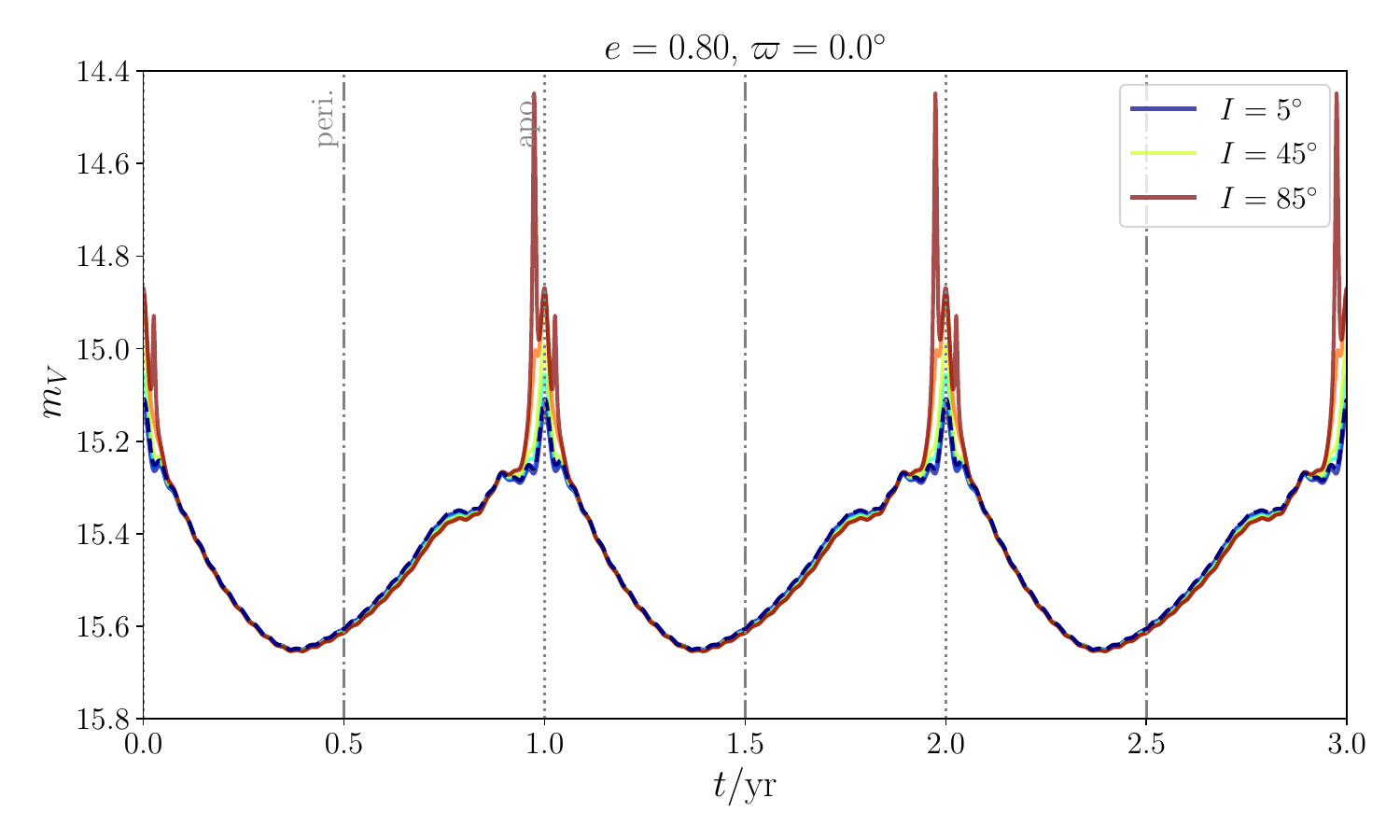} 
\end{array}$
\end{center}
\vspace{-15pt}
\caption{
The same as Figure \ref{Fig:LCs_pom} except each panel is for a different binary orbital eccentricity and the same viewing azimuth. Note that, as in Figure \ref{Fig:LCs_pom}, viewing angles $\varpi$ and $I$ affect only the Doppler boost and lensing signatures while the binary eccentricity affects both accretion variability and Doppler+lensing magnifications.
}
\label{Fig:LCs_ecc}
\end{figure*}

\paragraph{Example Lightcurves}
We compute example lightcurves in the V-band (optical) for different binary viewing angles and eccentricities in Figures \ref{Fig:LCs_pom} and \ref{Fig:LCs_ecc}. For these examples we choose binary parameters $M=2 \times 10^9 \Msun$, $P=1$yr, place the source at a luminosity distance of $1.5$~Gpc ($z\approx0.29$), and prescribe a total accretion rate onto the binary of $10\%$ of the Eddington rate, with $10\%$ accretion efficiency. To set the amplitude of Doppler-boost variability we choose a spectral index in the observing band of $\alpha=-1$ \citep[see, \eg,][]{Charisi+2018} and keep $\cos{I}=1$ fixed for easy comparison throughout.  For the accretion-rate ratio we use Eq. (\ref{Eq:Qfit}). For an eccentricity of $e=0.4$, this results in a value of $Q = 0.169$, and V-band flux ratios of $\chi_{V,1} = 0.308$ and $\chi_{V,2} = 0.160$. Hence, in this example, the minidisks are contributing $\approx 53\%$ of the total V-band flux. Note that this relative contribution can be a strong function of observing band, binary masses, and accretion rate, ranging from 0 to 1.
We compute the flux in the V-band, $F_V(t)$ by multiplying Eq.~(\ref{Eq:Magnfy}) by the sum of Eqs.~(\ref{Eq:Fnus}) evaluated at $\nu=5.5\times10^{14}$~Hz, and then computing an approximate V-band apparent magnitude $m_V = -2.5 \log_{10}\left[ F_V(t)/F_{V,0} \right]$, using the V-band zero-point flux of $F_{V,0}=3630.22$~Jy.

%%%%%%%%%%%%%%%%%%%%%%%%%%%%%%%%%%%%%%%%%%%%%%%%
%%% Lightcurves %%%
%%%%%%%%%%%%%%%%%%%%%%%%%%%%%%%%%%%%%%%%%%%%%%%%
%%%%%%%%%%%%%%%%%%%%%%%%%%%%%%%%%%%%%%%%%%%%%%%%

%%%%%%%%%%%%%%%%%%%%%%%%%%%%%%%%%%%%%%%%%%%%%%%%
%%% Lightcurves %%%
%%%%%%%%%%%%%%%%%%%%%%%%%%%%%%%%%%%%%%%%%%%%%%%%
%%%%%%%%%%%%%%%%%%%%%%%%%%%%%%%%%%%%%%%%%%%%%%%%

Figures \ref{Fig:LCs_pom} and \ref{Fig:LCs_ecc} show that a wide range of lightcurve morphologies arise when allowing viewing angle and eccentricity to vary. It is significant that each lightcurve is uniquely fixed by binary and observer parameters. Specifically, the Doppler+lensing-induced and accretion-rate-induced features cannot be shifted in phase independently of one another. This is because both accretion-rate and the Doppler+lensing variability have features fixed to specific values of the binary phase. The peaks and troughs of the Doppler boost modulation (as well as its shape) and the lensing flares occur at unique values of the binary phase for a given eccentricity and observing angle, while accretion-rate variability for eccentric orbits encodes the binary phase via accretion-rate peaks that occur near pericenter for prograde orbits and between pericenter and apocenter for retrograde orbits (see Figure \ref{Fig:Mdot_prog_ALL}). 

In contrast, for a near-circular-orbit binary, the peak of orbital timescale variability is related to the passage of a binary component by the near-side of the lopsided circumbinary cavity \citep[\eg,][]{DHM:2013:MNRAS} and so depends on the relative orientation of the cavity and not the binary phase with respect to the observer's line of sight. Put another way, a lightcurve exhibiting hydrodynamic and Doppler+lensing variability for a near-circular orbit binary would only allow identification up to an undetermined orientation of the circumbinary disk cavity on the sky. For an eccentric binary this free parameter is eliminated, and the lightcurve model is fully specified by binary parameters and the observer's viewing angle.

That the lightcurves in Figures \ref{Fig:LCs_pom} and \ref{Fig:LCs_ecc} are unique to the chosen binary parameters and observer angles offers considerable constraining power compared to the circular-orbit case. It also makes this signature more difficult to duplicate via non-SMBHB drivers of variability.

Finally, we note that for equal-mass binaries it is often assumed that the orbital Doppler effect is nullified or greatly diminished since both black-hole minidisks are assumed to be emitting at the same luminosity, and via the two-body problem, will have opposite line-of-sight velocities \citep[\eg,][]{PG1302Nature:2015b}. However, because the accretion rate can be split unequally between the components of eccentric binaries (Fig.~\ref{Fig:Qrat}), the Doppler boost can still cause significant modulations for equal mass binaries when the orbital eccentricity is non-zero.

\section{Discussion and Conclusion}
\label{S:Conclusion}

We have analysed the variability (Fig.~\ref{Fig:2DPow}) and relative magnitudes (Fig.~\ref{Fig:Qrat}) of accretion rates measured from viscous hydrodynamical simulations of disks accreting onto equal-mass, eccentric binaries in both prograde and retrograde configurations. With the goal of generating lightcurve models to facilitate searches for accreting binaries, we developed a tool, named \codename, which can rapidly generate accretion-rate time series at any eccentricity in our continuous sweep of simulations ($e\leq0.8$). We then post-processed these accretion-rate time series to generate simple models for lightcurves at optical wavelengths, including also observer-dependent effects of orbital Doppler boosting and gravitational self-lensing.

It is important to note that the details of the accretion rates presented here will likely vary with different included physics, sink prescriptions, and numerical methods for solving the equations of hydrodynamics. However, while these simulations correspond to a simplest non-trivial inclusion of 2D, viscous, isothermal hydrodynamics, they capture some robust features that lead to accretion-rate periodicities observed over a wide range of calculations that include different physics: 3D \citep[\eg,][]{Moody:2019}, self-gravity \citep{Franchini_eccSG+2024}, magneto-hydrodynamics \citep[MHD, \eg,][]{ShiKrolik:2012:ApJ, ShiKrolik:2015}, General Relativity \citep[\eg,][]{Noble+2021}, non-isothermal equations of state  
\citep[\eg,][]{Westernacher-Schneider:2022, WangBaiLaiII:2023}, for fixed and live binaries \citep[\eg,][]{Franchini_LiveBin+2023}, and are robust over a wide range of numerical techniques \citep{KITP_CC+2024}. Hence, while exact shapes of accretion-rate times series will depend on the physical parameters and numerical methods employed, the accretion-rate times series available through \codename~can give insight into the types of accretion variability expected and aid in building templates with which to search for such signatures.

Furthermore, the simple lightcurve models presented in Section \ref{S:Application}, could be expanded and adapted to numerous situations. More complex spectra that take into account different accretion flow properties could be added to this picture, \eg, emission characteristics of radiatively inefficient accretion flows, \citep[see the Methods Section of][]{PG1302Nature:2015b}, or those tailored to proto-planetary disks \citep[\eg,][]{Zhu_accCPD:2015}. Timescales for the disk spectra to respond to the changing accretion rate could also be taken into account. For example, the lightcurve generation procedure presented here could be modified by smoothing the reconstructed accretion rates in time with a smoothing kernel set by a buffering timescale due to, \eg, photon diffusion.
Beyond this, the fluid properties of the disk (via post-processing or inclusion of radiative cooling terms in the energy equation) can be used to generate mock spectra, as has been done for a much smaller parameter space in a number of works using viscous hydrodynamics \citep[\eg,][]{Farris:2015:Cool, Yike:2018, Westernacher-Schneider:2022, Krauth+2023, Franchini_decouple+2024, Cocchiararo+2024}, as well as general relativistic MHD \citep[\eg][]{dAscoli+2018, Combi:2021, Gutierrez+2022, Avara+2023}.

In addition to advancing lightcurve models with the accretion-rate time series investigated here, the accretion rates accessible with this tool should also be updated with the newest, and a wider range of, simulation results. Utilising both simple and fast simulations, which will expand available data to a wider range of parameter space (\eg, a wider range of binary and disk parameters for the types of simulations analysed here \citep{D'Orazio:CBDTrans:2016, Tiede:2020, Dittmann:2022, DittmannRyan:2023}), and also simulations including more physics that can improve accuracy in smaller portions of parameter space. We plan to add such improvements over time from our own calculations and also from the wider community. \\

\section{Public availability : \codename}

We have developed a simple Python package for rapidly generating periodic accretion rate time series and associated flux series at any eccentricity in our continuous sweep simulations. 

\codename\, is available in the Python Package Index, and it can be installed with
\begin{lstlisting}
    python -m pip install binlite
\end{lstlisting}
and imported locally as
\begin{lstlisting}
    import binlite as blt
\end{lstlisting}
It contains two main modules
\begin{lstlisting}
    blt.accretion
    blt.flux
\end{lstlisting}
for generating variability series of the mass accretion onto the binary and the flux at a given frequency (under the assumptions detailed in Section~\ref{S:Application}) respectively.
The source code and more detailed documentation are also available at \href{https://github.com/nbia-gwastro/binlite/}{github.com/nbia-gwastro/binlite}.

\acknowledgements
D.J.D. received funding from the European Union's Horizon 2020 research and innovation programme under Marie Sklodowska-Curie grant agreement No. 101029157. D.J.D. and C.T. acknowledge support from the Danish Independent Research Fund through Sapere Aude Starting Grant No. 121587.  P.D. acknowledges support from the National Science Foundation under grant AAG-2206299.

\bibliographystyle{apj} 
\bibliography{refs}
\end{document}